\documentclass[prd,aps,twocolumn,amsmath,amssymb,nofootinbib,preprintnumbers]
{revtex4}

\usepackage{graphicx}
\usepackage{dcolumn}
\usepackage{bm}
\usepackage{amsmath}
\usepackage{amsthm}
\usepackage{amsfonts}
\usepackage{euscript}
\usepackage{ifthen}
\usepackage{psfrag}
\usepackage{slashed}

\def\ls{\mathrel{\lower4pt\vbox{\lineskip=0pt\baselineskip=0pt
           \hbox{$<$}\hbox{$\sim$}}}}
\def\gs{\mathrel{\lower4pt\vbox{\lineskip=0pt\baselineskip=0pt
           \hbox{$>$}\hbox{$\sim$}}}}
\def\drawbox#1#2{\hrule height#2pt
\hbox{\vrule width#2pt height#1pt \kern#1pt
              \vrule width#2pt}
              \hrule height#2pt}

\def\Asym#1#2{\vcenter{\vbox{\drawbox{#1}{#2}
              \kern-#2pt
              \drawbox{#1}{#2}}}}

\def\nn{\nonumber}

\newcommand{\be}{\begin{equation}}
\newcommand{\ee}{\end{equation}}
\newcommand{\bea}{\begin{eqnarray}}
\newcommand{\eea}{\end{eqnarray}}

\newcommand{\gsim}{\lower.7ex\hbox{$\;\stackrel{\textstyle>}{\sim}\;$}}
\newcommand{\lsim}{\lower.7ex\hbox{$\;\stackrel{\textstyle<}{\sim}\;$}}

\newcommand{\vo}{\mathcal{V}}

\newcommand{\ben}{\begin{enumerate}}
\newcommand{\een}{\end{enumerate}}
\newcommand{\bei}{\begin{itemize}}
\newcommand{\eei}{\end{itemize}}
\newcommand{\bi}{\begin{itemize}}
\newcommand{\ei}{\end{itemize}}
\newcommand{\mc}{\mathcal}

\begin{document}

\title{Superheavy Dark Matter from String Theory}

\author{Rouzbeh Allahverdi$^{1}$}
\author{Igor Br\"ockel$^{2,3}$}
\author{Michele Cicoli$^{2,3}$}
\author{Jacek K. Osi\'nski$^{1}$}

\affiliation{$^{1}$ Department of Physics and Astronomy, University of New Mexico, Albuquerque, NM 87131, USA \\
$^{2}$ Dipartimento di Fisica ed Astronomia, Universit\`a di Bologna, via Irnerio 46, 40126 Bologna, Italy \\
$^{3}$ INFN, Sezione di Bologna, viale Berti Pichat 6/2, 40127 Bologna, Italy}

\begin{abstract}
Explicit string models which can realize inflation and low-energy supersymmetry are notoriously difficult to achieve. Given that sequestering requires very specific configurations, supersymmetric particles are in general expected to be very heavy implying that
the neutralino dark matter should be overproduced in a standard thermal history. However, in this paper we point out that this is generically not the case since early matter domination driven by string moduli can dilute the dark matter abundance down to the observed value. We argue that generic features of string compactifications, namely a high supersymmetry breaking scale and late time epochs of modulus domination, might imply superheavy neutralino dark matter with mass around $10^{10}-10^{11}$ GeV. Interestingly, this is the right range to explain the recent detection of ultra-high-energy neutrinos by IceCube and ANITA via dark matter decay.
\end{abstract}

\maketitle

\section{Introduction}

While there are various lines of evidence for the existence of dark matter (DM) in the universe~\cite{BHS}, the nature of DM remains a major problem at the interface of cosmology and particle physics. Weakly interacting massive particles (WIMPs) have long been a promising candidate and the focus of most direct, indirect and collider searches. In an attractive scenario, called the `WIMP miracle', the DM relic abundance is obtained via thermal freeze-out in a radiation dominated (RD) universe for the nominal value of the DM annihilation rate $\langle \sigma_{\rm ann} v \rangle = 3 \times 10^{-26}$ cm$^3$ s$^{-1}$. This scenario, however, has been coming under increasing scrutiny by recent experiments, namely the Fermi-LAT results from observations of dwarf spheroidal galaxies~\cite{Fermi1} and newly discovered Milky Way satellites~\cite{Fermi2}. A recent analysis~\cite{Beacom} has specifically ruled out thermal DM with a mass below 20 GeV in a model-independent way (unless there is P-wave annihilation or co-annihilation). Masses up to 100 GeV can be excluded if specific annihilation channels are considered.

The situation is different if the universe is not RD at the time of DM freeze-out~\cite{KT}. This typically happens in non-standard thermal histories where the universe is not in a RD phase from inflationary reheating all the way to Big Bang nucleosynthesis (BBN)~\cite{Review}. An important example is an epoch of early matter domination (EMD) driven by a component whose equation of state is the same as matter. This is a generic feature of early universe models arising from string theory constructions~\cite{KSW, Bobby, Cicoli:2016olq}. In this context, a string modulus is displaced from the minimum of its potential during inflation. Due to its long lifetime, the modulus dominates the energy density and gives rise to a period of EMD in the post-inflationary history. The modulus eventually decays and a RD universe is established prior to BBN. Various production mechanisms during EMD can yield the correct DM abundance for both $\langle \sigma_{\rm ann} v \rangle < 3 \times 10^{-26}$ cm$^3$ s$^{-1}$ and $\langle \sigma_{\rm ann} v \rangle > 3 \times 10^{-26}$ cm$^3$ s$^{-1}$~\cite{Howie}. 

Furthermore, the DM relic abundance can be completely decoupled from $\langle \sigma_{\rm ann} v \rangle$ if its main source is direct production from the decay of the component that drives an EMD phase~\cite{GG}. In this scenario, the relic abundance depends on the branching fraction for decay to DM (hence the `branching scenario'~\cite{Clado}) and the yield from the decay of the matter-like component. Non-thermal production of supersymmetric DM via the branching scenario has been studied in explicit string theory constructions where the volume modulus drives an epoch of EMD just before the onset of BBN~\cite{Allahverdi:2013noa}. A successful realization along these lines seems to be challenging for two reasons. First, the branching fraction of the volume modulus to DM is such that the correct abundance can be obtained for DM $\sim {\cal O}(10)$ GeV. Second, the decay of the volume modulus typically produces dark radiation (DR) in addition to DM, and avoiding an excess of DR severely constrains the branching scenario~\cite{Allahverdi:2014ppa}. 

However, in this work we shall show that the branching scenario could instead arise very generically in 4D string models with superheavy WIMPs. Several scenarios of supersymmetry breaking and inflation have already been realized in the context of string theory. Combining low-energy supersymmetry with successful inflationary model building is notoriously hard to achieve \cite{Kallosh:2004yh}. The main reason is that the requirement of obtaining density perturbations of the correct size tends to fix the inflationary scale at relatively high energies. In turn, masses of the supersymmetric particles are also generically pushed to large values, typically at an intermediate scale around $10^{10}-10^{11}$ GeV. A possible way to reconcile inflation with low-energy supersymmetry is to sequester the visible sector from the source of supersymmetry breaking in the bulk of the extra dimensions. Sequestered models, however, require a very specific brane configuration and K\"ahler metric for matter fields \cite{sequestering, Aparicio:2014wxa}. This solution therefore is not very generic. We note that statistical studies showed that high scale supersymmetry is a generic feature of the string landscape regardless of inflation \cite{Denef:2004ze, Broeckel:2020fdz}. 

Though the thermal DM scenario is known to overproduce superheavy WIMPs~\cite{unitarity}, the DM abundance may be diluted by epochs of EMD driven by string moduli. Hence two generic features of string compactifications, high-scale supersymmetry breaking and late time epochs of modulus domination, can successfully accommodate superheavy WIMPs with a mass around $10^{10}-10^{11}$ GeV. Incidentally, if such a DM candidate is unstable and has the right coupling to neutrinos, its decay into very energetic neutrinos could provide a tantalizing explanation of the ultra-high-energy cosmic rays recently observed by IceCube and ANITA \cite{Heurtier:2019git}.

We will illustrate this general picture by presenting an explicit model that involves two periods of EMD. The first one is driven by inflaton oscillations at the end of which the inflaton mainly decays to DR in a hidden sector, and produces superheavy DM via its tiny coupling to the visible sector. A second stage of EMD is driven by the volume modulus, which is dominantly coupled to the visible sector and is lighter than the DM. As a result, this second EMD phase only dilutes the abundance of DM and DR produced in inflaton decay down to observationally acceptable values. 

This paper is organized as follows. In Sec. \ref{Sec2} we briefly review the branching scenario for DM production. In Sec. \ref{Sec3} we discuss a successful framework for production of superheavy DM via the branching scenario. In Sec. \ref{Sec4} we introduce an explicit string theory model for realizing this scenario. In Sec. \ref{Sec5} we identify the allowed parameter space of this model for a successful inflation and a correct DM abundance, and we present numerical results for the post-inflationary evolution for a benchmark point. We conclude in Sec. \ref{Sec6}, and discuss generalized scenarios that involve more than one modulus in App. \ref{App}.

\section{Branching scenario: a brief review}
\label{Sec2}

Let us consider a post-inflationary history that includes an EMD era driven by coherent oscillations or non-relativistic quanta of a long-lived scalar field $\varphi$ with mass $m_\varphi$ and decay width $\Gamma_\varphi$. The continuous decay of $\varphi$ feeds radiation (assuming that decay products thermalize immediately) during the period that it dominates the energy density of the universe. The decay of $\varphi$ completes when the Hubble expansion rate is $H \simeq \Gamma_\varphi$, at which time the universe enters a RD phase. The resulting reheat temperature is $T_{\rm R} = (90/\pi^2 g_{*,{\rm R}})^{1/4} (\Gamma_\varphi M_{\rm P})^{1/2}$, where $g_{*,{\rm R}}$ is the number of relativistic degrees of freedom at $T = T_{\rm R}$.

The energy densities of $\varphi$ and radiation, denoted by $\rho_\varphi$ and $\rho_{\rm R}$ respectively, and the number density $n_\chi$ of DM particles $\chi$ are found by solving the following system of Boltzmann equations:
\bea
\label{boltzmann}
&&{\dot \rho_{\rm R}} + 4H\rho_{\rm R} = \Gamma_\varphi\rho_\varphi \, , \nn \\
&&{\dot \rho_\varphi} + 3H\rho_\varphi = - \Gamma_\varphi\rho_\varphi \, ,  \\
&&{\dot n}_\chi + 3 H  n_\chi =  \langle \sigma_{\rm ann} v \rangle  \left(n^2_{{\rm\chi, eq}} - n^2_\chi \right) + {\rm Br}_\chi \Gamma_\varphi  n_\varphi \, .  \nn
\eea
The first term on the right-hand side (RHS) of the last equation accounts for DM annihilation and inverse annihilation from the thermal bath ($\langle \sigma_{\rm ann} v \rangle$ denotes the thermally-averaged annihilation/inverse annihilation rate). The second term accounts for direct production of DM from $\varphi$ decay~\cite{Direct} (where ${\rm Br}_\chi$ is the number of DM particles produced per $\varphi$ decay). Freeze-out/in of DM happens during the EMD epoch if $T_{\rm f} > T_{\rm R}$, where $T_{\rm f} \simeq m_\chi/20$ in the case of freeze-out and $T_{\rm f} \simeq m_\chi/4$ for freeze-in~\cite{GKR:2000ex,E}.

Assuming that freeze-out/in production is subdominant, the main contribution to the DM relic density comes from direct production at $H \simeq \Gamma_\varphi$, and the number density of DM particles at this time is given by:
\be
\label{DMdens1}
n_\chi \simeq {\rm Br}_\chi n_\varphi = \frac{3 \Gamma^2_\varphi M^2_{\rm P}}{m_\varphi} ~ {\rm Br}_\chi\,.
\ee
The comoving number density of DM follows this expression, hence the name `branching scenario'~\cite{GG,Clado}, provided that residual annihilation of DM particles to the thermal bath is inefficient. This will be the case if $\langle \sigma_{\rm ann} v \rangle n_\chi < \Gamma_\varphi$, where $n_\chi$ is substituted from ~(\ref{DMdens1}). Otherwise, partial annihilation will somewhat reduce the DM number density leading to the so-called `annihilation scenario' of DM production~\cite{Ann1,Ann2}. The annihilation scenario can only be successful if $\langle \sigma_{\rm ann} v \rangle > 3 \times 10^{-26}$ cm$^3$ s$^{-1}$, which happens to be the case for weak-scale Wino and Higgsino DM. For small values of $\langle \sigma_{\rm ann} v \rangle$, as in the case of Bino DM or for $m_\chi \gtrsim 100$ TeV, only the branching scenario can yield the correct DM abundance.

After normalizing $n_\chi$ in (\ref{DMdens1}) by the entropy density $s = 2 \pi^2 g_{*,{\rm R}} T^3_{\rm R}/45$ at $T = T_{\rm R}$, the DM relic abundance in the branching scenario is found to be:
\be
\label{DMdens2}
\frac{n_\chi}{s} = \frac{3 T_{\rm R}}{4 m_\varphi} ~ {\rm Br}_\chi\,.
\ee
Here $3 T_{\rm R}/4 m_\varphi$ is the yield factor that is related to dilution due to entropy released by $\varphi$ decay. In order for the branching scenario to work, this must match the observed value:
\be
\label{DMobsdens}
\left(\frac{n_\chi}{s}\right)_{\rm obs} \simeq 4.2 \times 10^{-10}\left(\frac{1\,{\rm GeV}}{m_\chi}\right)\,.
\ee

A natural question is if the branching scenario can be successfully realized in explicit particle physics models of the early universe. This issue has been discussed in the context of type IIB string compactifications where $\varphi$ is the volume modulus~\cite{Allahverdi:2013noa}. In this case, we have $T_{\rm R}/m_\varphi \simeq (m_\varphi/M_{\rm P})^{1/2}$. Also, for supersymmetric DM, three-body decays of $\varphi$ result in a lower bound ${\rm Br}_\chi \gtrsim {\cal O}(10^{-3})$. Considering that $T_{\rm R} \gtrsim 3$ MeV (corresponding to $m_\varphi \gtrsim 50$ TeV) is required for BBN, (\ref{DMdens2}) and (\ref{DMobsdens}) imply that the correct DM abundance can be obtained for $m_\chi \lesssim {\cal O}(10)$ GeV. Moreover, avoiding excessive production of DR, which typically accompanies DM production in string compactifications \cite{Cicoli:2012aq, WinoDMDR,Cicoli:2015bpq,Cicoli:2018cgu}, seems to favor the annihilation scenario~\cite{Allahverdi:2014ppa}.

\section{Branching scenario and superheavy DM}
\label{Sec3}

In this Section we lay down a framework for production of superheavy DM via the branching scenario. To overcome the challenges mentioned in Sec. \ref{Sec2}, we invoke two epochs of EMD driven by the inflaton and a modulus field respectively, as in generic string models. We also consider constraints from the cosmic microwave background (CMB) on such a scenario. In Sec. \ref{Sec4} we shall present an explicit type IIB string model that successfully realizes this scenario (see also App. \ref{App} for another explicit string model which realizes this scenario with an additional epoch of moduli domination).

\subsection{Scenario with an epoch of modulus domination}
\label{Sec3a}

The scenario we consider involves two periods of EMD driven by the inflaton $\sigma$ and a modulus field $\phi$ in succession. Both of these fields behave as the field $\varphi$ described in Sec. \ref{Sec2}. The inflaton $\sigma$ is responsible for inflation at the end of which the Hubble expansion rate is $H_{\rm inf}$. The inflaton mass at the minimum of its potential is $m_\sigma$ and its couplings to the visible and hidden sectors are $c_{\rm vis}/M_{\rm P}$ and $c_{\rm hid}/M_{\rm P}$ respectively where $c_{\rm vis} \ll c_{\rm hid}$. We will also assume that there is no stable non-relativistic particle in the hidden sector, so that the inflaton decay into hidden sector degrees of freedom just produces DR. Therefore, the inflaton decay rate into DR dominates over the one into visible sector particles since $\Gamma_{\sigma\to{\rm DR}} \simeq c^2_{\rm hid} m^3_\sigma/M^2_{\rm P} \gg \Gamma_{\sigma\to{\rm vis}}\simeq c^2_{\rm vis} m^3_\sigma/M^2_{\rm P}$. We will also denote the total inflaton decay width as $\Gamma_\sigma = \Gamma_{\sigma\to{\rm vis}} + \Gamma_{\sigma\to{\rm DR}}$.

The modulus $\phi$ has mass $m_\phi < m_\sigma$. Its coupling to the visible sector is $d_{\rm vis}/M_{\rm P}$, while its coupling to the hidden sector is $d_{\rm hid}/M_{\rm P}$ with $d_{\rm vis} \gg d_{\rm hid}$. We will assume again that the modulus decay into the hidden sector produces just DR. This gives $\Gamma_{\phi\to{\rm vis}} \simeq d^2_{\rm vis} m^3_\phi/M^2_{\rm P} \gg \Gamma_{\phi\to{\rm DR}}\simeq d^2_{\rm hid} m^3_\phi/M^2_{\rm P}$. The total modulus decay width is instead $\Gamma_\phi = \Gamma_{\phi\to{\rm vis}} + \Gamma_{\phi\to{\rm DR}}$. We assume that $m_\phi < m_\chi$ so that $\phi$ decay to DM is kinematically forbidden. The modulus acquires a displacement $\phi_0$ from the minimum of its potential during inflation.

Below, we summarize the important stages of the post-inflationary history in this scenario in chronological order:
\vskip 1.5mm
\noindent
{\bf 1-} $\Gamma_\sigma \lesssim H < H_{\rm inf}$: The universe is in an EMD phase driven by inflaton oscillations about the minimum of its potential.
$\phi$ also starts oscillating at this stage and $\rho_\phi = (\phi_0/M_{\rm P})^2 \rho_\sigma$. The inflaton decay completes at $H \simeq \Gamma_\sigma$ and mainly populates the hidden sector.
\vskip 1.5mm
\noindent
{\bf 2-} $H_{\rm D} \lesssim H < \Gamma_\sigma$: The universe is in a RD phase at this stage. The modulus oscillations behave like matter, and hence $\rho_\phi$ is redshifted more slowly than $\rho_{\rm R}$. As a result, $\phi$ starts to dominate at $H_{\rm D} \simeq (\phi_0/M_{\rm P})^4 \Gamma_\sigma$, which is the onset of a second phase of EMD.
\vskip 1.5mm
\noindent
{\bf 3-} $\Gamma_\phi \lesssim H < H_{\rm D}$: The universe is in a modulus-driven EMD epoch during this stage. The modulus decay completes when the Hubble expansion rate is  $H \simeq \Gamma_\phi$ and reheats the visible sector. This results in the formation of a RD universe prior to the onset of BBN.
\vskip 1.5mm

The inflaton decay to the visible and hidden sectors produces DM and DR respectively. Given that $m_\chi > m_\phi$, the modulus decay dilutes both abundances and reproduces some amount of DR in the hidden sector. The number density of DM particles directly produced by the inflaton decay at $H \simeq \Gamma_\phi$ is:
\be
n_\chi \simeq n_\sigma ~ {\rm Br}_\chi ~ \left(\frac{a_\sigma}{a_{\rm D}}\right)^3 \left(\frac{a_{\rm D}}{a_\phi}\right)^3 \,,
\label{nDM1}
\ee
where $n_\sigma = 3 \Gamma^2_\sigma M^2_{\rm P}/m_\sigma$ is the inflaton number density at the end of stage 1, $(a_\sigma/a_{\rm D})^3 = (H_{\rm D}/\Gamma_\sigma)^{3/2}$ is the number density redshift during stage 2, and $(a_{\rm D}/a_\phi)^3 =  (\Gamma_\phi/{H_{\rm D}})^2$ is the number density redshift during stage 3.

If DM is the lightest $R$-parity odd particle in the visible sector, we have:
\be
{\rm Br}_\chi \simeq \frac{\Gamma_{\sigma\to{\rm vis}}}{\Gamma_\sigma}
~ {\rm Br}_{\rm vis,odd}\,.
\ee
The first factor on the RHS of this expression is the fraction of $\sigma$ quanta that decay to the visible sector. The second factor is the ratio of the number of $R$-parity odd particles (which subsequently decay to DM) to the total number of particles in the visible sector produced per $\sigma$ decay. In the explicit example that we discuss later, the $\sigma$ decay into the visible sector mainly occurs through two-body decays to gauge fields. Two-body decays to $R$-parity odd particles are highly suppressed, but they are produced via three-body decays including one gauge field and two gauginos resulting in ${\rm Br}_{\rm vis,odd}\simeq 10^{-3}$ (which is essentially a phase space factor)~\cite{Clado}.

Therefore, after normalizing $n_\chi$ by the entropy density $s$, we find:
\be
\frac{n_\chi}{s} \simeq \frac34 \times 10^{-3} ~ \frac{1}{Y_\phi^2}
~ \frac{\Gamma_{\sigma\to{\rm vis}}}{\Gamma_\sigma}
~ \frac{\Gamma_\phi}{\Gamma_{\phi\to{\rm vis}}}
~ \frac{T_{\rm R}}{m_\sigma} \,,
\label{DMrelab1}
\ee
where:
\be
T_{\rm R} = \left(\frac{90}{\pi^2 g_{*,{\rm R}}}\,\frac{\Gamma_{\phi\to{\rm vis}}}{\Gamma_\phi}\right)^{1/4} \sqrt{\Gamma_\phi M_{\rm P}}\,, 
\label{TR}
\ee
with $g_{*,{\rm R}}$ denoting the number of relativistic degrees of freedom in the visible sector at $T = T_{\rm R}$, and $Y_\phi \equiv \phi_0/M_{\rm P}$.

Regarding DR, its energy density at $H \simeq \Gamma_\phi$ is:
\be
\rho_{\rm DR} \simeq \rho_\sigma\,\frac{\Gamma_{\sigma\to{\rm DR}}}{\Gamma_\sigma} \left({a_\sigma \over a_{\rm D}}\right)^4 \left({a_{\rm D} \over a_\phi}\right)^4 + \rho_\phi\,\frac{\Gamma_{\phi\to{\rm DR}}}{\Gamma_\phi}\,,
\label{DR1}
\ee
where $\rho_\sigma \simeq 3 \Gamma^2_\sigma M^2_{\rm P}$, $(a_\sigma/a_{\rm D})^4 = (H_{\rm D}/\Gamma_\sigma)^2$ is the energy density redshift during stage 2, $(a_{\rm D}/a_\phi)^4 = (\Gamma_\phi/H_{\rm D})^{8/3}$ is the energy density redshift during stage 3, and $\rho_\phi \simeq 3 \Gamma^2_\phi M^2_{\rm P}$. Hence, the final fractional energy density of DR is given by:
\be
{\rho_{\rm DR} \over \rho_{\rm R}}
\simeq \frac{1}{Y_\phi^{8/3}} \left(\frac{\Gamma_\phi}{\Gamma_\sigma}\right)^{2/3}\,\frac{\Gamma_{\sigma\to{\rm DR}}}{\Gamma_\sigma}\,\frac{\Gamma_\phi}{\Gamma_{\phi\to{\rm vis}}}+\frac{\Gamma_{\phi\to{\rm DR}}}{\Gamma_{\phi\to{\rm vis}}}\,.
\label{DRrelab1}
\ee
This ratio must be small enough to satisfy the observational constraints on the DR abundance.

\subsection{Constraints from CMB}

Inflation is the dominant paradigm for generating the almost scale-invariant perturbations. The number of e-foldings between the time when perturbations of a given wavelength exit the horizon and the end of inflation depends on the scale of inflation as well as the post-inflationary thermal history. One or more periods of EMD change the number of e-foldings from that in a standard thermal history.

In the scenario discussed in Sec. \ref{Sec3a}, the number of e-foldings of inflation between the time when the pivot scale $k_* = 0.05$ Mpc$^{-1}$ left the horizon and the end of inflation can be written as~\cite{LL, RG}:
\be
N_{\rm e} \simeq 57 + {1 \over 4} \ln r - {1 \over 4} N_{\rm reh} - {1 \over 4} N_{\phi} ,
\ee
where $r$ is the tensor-to-scalar ratio and:
\bea
N_{\rm reh} &\simeq& {2 \over 3} \ln \left({H_{\rm inf} \over \Gamma_\sigma}\right) \, , \nn  \\
N_{\phi} &\simeq& {2 \over 3} \ln \left({H_{{\rm D}} \over \Gamma_{\phi}}\right) \simeq {2 \over 3} \ln \left(Y^4_{\phi} {\Gamma_\sigma \over \Gamma_{\phi}}\right) \, .
\eea
Here $N_{\rm reh}$ and $N_\phi$ denote the duration of EMD phases from inflationary reheating and modulus domination (stages 1 and 3 above) respectively. This results in:
\be \label{1modulus}
N_{\rm e} \simeq 57 + {1 \over 4} \ln r - {1 \over 6} \ln \left(Y^4_\phi\, {H_{\rm inf} \over \Gamma_\phi}\right) .
\ee

In important universality classes of inflation, the scalar spectral index $n_{\rm s}$ is related to $N_{\rm e}$ through a simple relation~\cite{Roest}:
\be
n_{\rm s} = 1 - {a \over N_{\rm e}}\,.
\ee
For example, in the Starobinsky model and Higgs inflation, as well as the specific model of string inflation that we will discuss later, $a = 2$. This then leads to:
\be
N_{\rm e} = {2 \over 1 - n_{\rm s}}\,.
\ee
This implies that:
\be
N_{\rm e} \gtrsim {2 \over 1 - n_{\rm {s,min}}},
\ee
where $n_{\rm {s,min}}$ is the minimum value in the 2$\sigma$ region allowed by Planck data~\cite{Planck}. For a given model of inflation where $H_{\rm inf}$ is known, this in turn sets an upper bound on $Y^4_\phi \Gamma^{-1}_\phi$
through (\ref{1modulus}).

On the other hand, for known inflaton parameters $m_\sigma$ and $\Gamma_\sigma$, (\ref{DMrelab1}) and (\ref{DRrelab1}) result in a lower bound on $Y^4_\phi \Gamma^{-1}_\phi$ in order not to overproduce DM and DR in our scenario.

Therefore, obtaining the correct abundance of DM (while avoiding an excessive production of DR) and getting an acceptable value of $n_{\rm s}$ constrain the epoch of modulus domination in opposite ways.\footnote{The implications of CMB constraints for non-thermal DM in low-scale supersymmetry has been studied in~\cite{ADM}.} This can be understood intuitively as follows. While diluting the abundance of DM and DR produced from inflaton decay to acceptable levels requires a long enough bout of modulus domination, satisfying the lower bound on $n_{\rm s}$ limits the duration of that period from above.

\section{A string model with an epoch of modulus domination}
\label{Sec4}

In this Section we shall present an explicit string model which successfully realizes inflation and superheavy DM via the branching scenario with an epoch of modulus domination.

\subsection{The setup}

We consider a type IIB model with 3 K\"ahler moduli $T_i=\tau_i + {\rm i} c_i$, $i=1,...,3$ and a Calabi-Yau volume of the form:
\be
\vo = \tau_{\rm big}^{3/2} - \tau_{\rm vis}^{3/2}-\tau_{\rm inf}^{3/2}\,.
\ee
The visible sector is realized via a stack of D7-branes wrapped around the 4-cycle whose volume is controlled by $\tau_{\rm vis}$, while inflation is driven by the modulus $\tau_{\rm inf}$ as in K\"ahler moduli inflation \cite{Conlon:2005jm}. A hidden sector lives instead on a stack of D7-branes wrapped around the 4-cycle whose volume is given by $\tau_{\rm inf}$.

The structure of the effective supergravity theory is determined by the K\"ahler potential $K$ and the superpotential $W$. $K$ is given by:
\be
K = -2\ln\left(\vo+\frac{\xi}{2 g_s^{3/2}}\right),
\ee
where $g_s$ is the string coupling and $\xi$ is an $\mathcal{O}(1)$ coefficient which controls $\alpha'$ corrections \cite{BBHL} beyond the tree-level expression. $W$ instead reads:
\be
W = W_0 + A_{\rm vis}\,e^{-a_{\rm vis} T_{\rm vis}} + A_{\rm inf}\,e^{-a_{\rm inf} T_{\rm inf}}\,,
\ee
where $W_0\sim\mathcal{O}(10-100)$ is the tree-level contribution, while the terms proportional to $A_{\rm vis}$ and $A_{\rm inf}$ are non-perturbative effects \cite{Blumenhagen:2009qh} (all $A$'s and $a$'s are expected to be $\mathcal{O}(1)$ constants).

Moduli stabilization produces a typical LVS minimum \cite{LVS} at exponentially large volume in string units, $\vo \simeq \tau_{\rm big}^{3/2}\sim e^{1/g_s}$, while the two blow-up modes are fixed at smaller values $\tau_{\rm vis}\sim \tau_{\rm inf}\sim 1/g_s \sim \mathcal{O}(10)$, where we take the string coupling in the perturbative regime $g_s\lesssim 0.1$. Notice that $\tau_{\rm vis}$ sets the value of the visible sector gauge coupling $\alpha_{\rm vis}^{-1}=4\pi g_{\rm vis}^{-2}= \tau_{\rm vis}\sim \mc{O}(10)$ which turns out to be in the appropriate phenomenological regime.

Moduli stabilization proceeds as follows: at leading order in a $1/\vo$ expansion, non-perturbative corrections to $W$ combined with $\alpha'$ corrections to $K$ stabilize $\vo$, $\tau_{\rm vis}$, $c_{\rm vis}$, $\tau_{\rm inf}$ and $c_{\rm inf}$, leaving 1 flat direction parameterized by the axion $c_{\rm big}$.\footnote{More precisely $\tau_{\rm vis}$ should be fixed by perturbative corrections to $K$ \cite{Cicoli:2008va} due to the interplay between chirality and non-perturbative effects \cite{Blumenhagen:2007sm}. However this detail is almost irrelevant for the phenomenological implications of our model.} This axion turns out to be ultra-light since it receives a tiny mass due to additional $T_{\rm big}$-dependent non-perturbative corrections to $W$. Thus $c_{\rm big}$ plays the role of hidden sector dark radiation. This system admits a non-supersymmetric AdS minimum which can however be uplifted to dS via several possible mechanisms (anti D3-branes \cite{KKLT}, T-branes \cite{Cicoli:2015ylx}, non-perturbative effects  at singularities \cite{Cicoli:2012fh}, non-zero F-terms of the complex structure moduli \cite{Gallego:2017dvd}).

\subsection{Moduli mass spectrum}

The determination of the moduli mass spectrum and couplings to both visible and hidden sector fields requires first to go to canonically normalized fields. Following the notation of Sec. \ref{Sec3}, we will denote them as: ($i$) $\sigma$ for $\tau_{\rm inf}$ since this modulus plays the role of the inflaton; ($ii$) $\phi$ for $\tau_{\rm big}$ since this modulus will give rise to an EMD epoch after the end of inflation; and ($iii$) $a_{\rm DR}$ for the closed string axion $c_{\rm big}$ which behaves as dark radiation. Defining:
\be
\epsilon \equiv \frac{W_0}{\vo}\ll 1  \qquad\text{and}\qquad \kappa \equiv \frac{g_s}{8\pi}\ll 1\,,
\ee
the mass spectrum of the relevant moduli around the minimum becomes \cite{Cicoli:2010ha,Cicoli:2010yj} (see \cite{Burgess:2010bz} for the correct normalization factor $\kappa$):
\bea
m_\sigma^2 &\simeq& \kappa \,\epsilon^2 \left(\ln\epsilon\right)^2 \, M_{\rm P}^2\nn \\
m_\phi^2 &\simeq& \frac{\epsilon\,m_\sigma^2}{g_s^{3/2} W_0\, |\left(\ln\epsilon\right)^3|}\ll m_\sigma^2\quad\text{for}\quad \epsilon \ll 1 \nn \\
m^2_{a_{\rm DR}} &\simeq& \kappa \,e^{-2 \vo^{2/3}}\,M_{\rm P}^2 \sim 0\,.
\label{ModMassSpectrum}
\eea

This setup allows to realize K\"ahler moduli inflation \cite{Conlon:2005jm} where the inflaton is $\sigma$ since this modulus becomes much lighter than $H \simeq m_\phi$ as soon as it is displaced from its minimum. $\tau_{\rm vis}$, $c_{\rm vis}$, and $c_{\rm inf}$ are heavy spectator fields which do not get displaced during inflation since their mass is of the same order of the mass of $\sigma$ around the minimum, and so it is much larger than $H$. On the other hand, all the other moduli get displaced from their minimum during inflation. We shall focus just on $\phi$ since the axion $a_{\rm DR}$ remains almost massless and behaves as a source of extra dark radiation. We shall also denote the displacement of the canonically normalized light K\"ahler modulus as $\phi_0 = Y_{\phi}\, M_{\rm P}$. Explicit computations have shown that $Y_{\phi} \simeq 0.01 - 0.1$ \cite{Cicoli:2016olq}. Due to this displacement during inflation, $\phi$ gives rise to a period of modulus domination. Moreover supersymmetry is broken due to non-zero F-terms of the K\"ahler moduli which generate a gravitino mass $m_{3/2}$ together with gaugino and scalar masses of order \cite{Conlon:2006wz}:
\be
m_{3/2} = \sqrt{\kappa}\,\epsilon\,M_{\rm P}\,,\qquad m_0 \simeq M_{1/2} \simeq \frac{m_{3/2}}{|\ln\epsilon|}\,.
\label{GravSoft}
\ee
Taking the DM mass of the same order as the soft terms, $m_\chi \simeq m_0 \simeq M_{1/2}$, we realize that:
\be
m_\phi^2 \simeq \frac{\epsilon\,|\ln\epsilon|}{g_s^{3/2} W_0}\,m_\chi^2\ll m_\chi^2\quad\text{for}\quad\epsilon\ll 1\,,
\ee
which ensures that DM cannot be reproduced from the decay of the light modulus $\phi$.

Notice that in order to avoid any cosmological moduli problem, the mass of $\phi$ has to be $m_\phi\gtrsim\mathcal{O}(50)$ TeV. Using (\ref{ModMassSpectrum}) and setting $g_s\simeq 0.1$ and $1\lesssim W_0\lesssim 100$, this gives the bound $5 \times 10^{-9}- 10^{-8}\lesssim \epsilon \ll 1$ which, when translated in terms of the overall volume, becomes $1 \ll \vo \lesssim 10^8 - 5\times 10^9$. This, in turn, produces a scenario of superheavy DM since it sets a lower bound on the DM mass of order $m_\chi \gtrsim 10^{10}- 10^{11}$ GeV. As we shall see in the Sec. \ref{Sec5}, values of $\vo$ below $10^8 - 5\times 10^9$ are also required to generate, during inflation, the observed value of the amplitude of the density perturbations.

\subsection{Hidden sector configuration}
\label{HidSec}

Let us comment a bit more on the configuration of the hidden sector D7-stack wrapping $\tau_{\rm inf}$. This has to provide a non-perturbative contribution to the superpotential which generates the inflationary potential, and be such that the inflaton decay into the hidden sector produces just relativistic degrees of freedom without additional contributions to the DM abundance. This dark radiation component is subsequently diluted by the decay of the lightest modulus. If the hidden sector is a supersymmetric $SU(N_c)$ theory with $N_f$ flavors, it would confine if $N_f<N_c$. The corresponding scale of strong dynamics $\Lambda$ can be shown to be above the inflaton mass, $m_\sigma < \Lambda$ \cite{Cicoli:2010ha}, and so $\sigma$ cannot decay into glueballs ($gg$), `gluinoballs' ($\tilde{g}\tilde{g}$), and `glueballinos' ($g\tilde{g}$) since they all develop a mass of order $\Lambda$. Hence we need to discard the pure SYM case. For $N_f>0$ with soft supersymmetry breaking terms, squarks and quarks form scalar and fermionic condensates which all develop a mass of order $m_0\simeq M_{1/2} \ll m_\sigma$ \cite{Martin:1998yr}, except for pion-like mesons which are exactly massless in the absence of a supersymmetric quark mass term in $W$. Therefore $\sigma$ could decay into these heavy condensates but some of them would be stable in the absence of EW interactions. We conclude that the hidden sector cannot be a simple $SU(N_c)$ theory with $N_f$ flavors. The best configuration for the hidden sector is instead a copy of the visible sector, i.e. an MSSM-like hidden sector, with however 3 differences with respect to the visible sector: ($i$) the scale of strong dynamics $\Lambda$ is much higher than in ordinary QCD; ($ii$) $R$-parity is not conserved so that hidden protons are unstable; ($iii$) the mass of the hidden electrons is very small so that they are still relativistic, like neutrinos. In this scenario, all hidden degrees of freedom produced from the inflaton decay eventually decay into hidden massless gauge bosons or hidden relativistic matter fermions.

A final requirement is the absence of any leakage of energy between hidden and visible sector degrees of freedom due to kinetic mixing between $U(1)$s or a possible moduli portal. The first option can be avoided by construction if the hidden gauge group does not contain any Abelian $U(1)$ factor.\footnote{Even in the presence of a $U(1)$ kinetic mixing, we expect the mixing parameter to be very small due to the geometric separation between $\tau_{\rm vis}$ and $\tau_{\rm hid}$ \cite{Cicoli:2011yh}.} On the other hand, a moduli portal between the two sectors could be created by the volume modulus $\phi$. However, we expect this effect to be negligible since, as we shall see in Sec. \ref{ModCoupl}, this field couples only with Planckian strength to both sectors, and so any leakage would be proportional to $(1/M_{\rm P})^4$.

\subsection{Moduli couplings and decay rates}
\label{ModCoupl}

Due to the geometric separation in the extra dimensions between $\tau_{\rm vis}$ (which supports the visible sector D7 stack) and $\tau_{\rm inf}$ (which supports a hidden sector D7 stack), the coupling of the canonically normalized inflaton $\sigma$ to hidden sector gauge bosons is much stronger than the one to visible sector gauge fields \cite{Cicoli:2010ha}:
\be
\mathcal{L}\supset -\frac14 \frac{c_{\rm hid}}{M_{\rm P}}\,\sigma\,F_{\mu\nu}^{\rm hid}F^{\mu\nu}_{\rm hid}
-\frac14 \frac{c_{\rm vis}}{M_{\rm P}}\,\sigma\,F_{\mu\nu}^{\rm vis}F^{\mu\nu}_{\rm vis}\,,
\label{Linflcoupl}
\ee
with:
\be
c_{\rm hid} \simeq g_s^{3/4}\,\sqrt{\vo}\gg 1\qquad\text{and}\qquad c_{\rm vis} \simeq c_{\rm hid}^{-1}\,.
\label{chidcvis}
\ee
Notice that the interactions in (\ref{Linflcoupl}) provide the main contributions to the inflaton decay rate to both visible and hidden degrees of freedom. In fact, since $m_0\simeq M_{1/2} \ll m_\sigma$, the inflaton decay into supersymmetric partners is mass suppressed. The same consideration applies to the inflaton decay into both visible and hidden sector matter fermions. The decay rate into Higgses is also mass suppressed except for the case of a Giudice-Masiero interaction in $K$ which we assume to be absent.\footnote{Including a Giudice-Masiero coupling between $\sigma$ and Higgs degrees of freedom would not modify our results qualitatively.} This implies the following important relation for the determination of the DM abundance using (\ref{DMrelab1}):
\be
\frac{\Gamma_{\sigma\to{\rm vis}}}{\Gamma_\sigma} = \frac{N_g}{N_g^{\rm hid}}\frac{1}{c_{\rm hid}^4}\frac{1}{\left(1+\frac{N_g}{N_g^{\rm hid}}\frac{1}{c_{\rm hid}^4}\right)}\simeq \frac{N_g}{N_g^{\rm hid}}\frac{1}{g_s^3\,\vo^2} \ll 1 \,,
\label{ratiocvischid}
\ee
where we included also the number of visible and hidden sector gauge bosons denoted respectively as $N_g$ and $N_g^{\rm hid}$. For an MSSM-like visible sector we have $N_g=12$ while $N_g^{\rm hid}$ is a model-dependent parameter which can also be larger than $N_g$.

On the other hand the light modulus $\phi$ can decay to:
\bi
\item Hidden sector gauge bosons:
\be
\mathcal{L}\supset -\frac14 \frac{\lambda_{\rm hid}}{M_{\rm P}}\,\phi\,F_{\mu\nu}^{\rm hid}F^{\mu\nu}_{\rm hid}\,,
\qquad \lambda_{\rm hid}\simeq \frac{1}{|\ln\epsilon|} \nn
\ee

\item Dark radiation bulk axions:
\be
\mathcal{L}\supset \lambda_{\rm DR}\, \frac{m_\phi^2}{M_{\rm P}}\,\phi\,a_{\rm DR}\,a_{\rm DR}\,,
\qquad \lambda_{\rm DR}\simeq \frac{1}{\sqrt{6}} \nn
\ee

\item Visible sector gauge bosons:
\be
\mathcal{L}\supset -\frac14 \frac{\lambda_{\rm vis}}{M_{\rm P}}\,\phi\,F_{\mu\nu}^{\rm vis}F^{\mu\nu}_{\rm vis}\,,
\qquad \lambda_{\rm vis}\simeq \frac{1}{|\ln\epsilon|} \nn
\ee

\item Visible sector Higgs $h^0$ and would-be Goldstone bosons $G^0$ and $G^\pm$ \cite{Cicoli:2015bpq}:
\be
\mathcal{L}\supset c\,\frac{m_\phi^2}{M_{\rm P}}\,\phi\left[(h^0)^2+(G^0)^2+({\rm Re} G^+)^2+({\rm Im} G^+)^2\right] \nn
\ee
with $c = Z/(2\sqrt{6})$ where $Z$ is an $\mathcal{O}(1)$ parameter controlling Giudice-Masiero contributions to the K\"ahler potential of the form $K\supset \frac{Z}{\tau_b} (H_u H_d+{\rm h.c.})$ \cite{Cicoli:2012aq}.
\ei

For $\epsilon \ll 1$, the decay rate of $\phi$ into both hidden and visible gauge bosons is suppressed. On the other hand, the decay of $\phi$ into ultra-light bulk axions could give rise to extra dark radiation which needs to be in agreement with present observational bounds \cite{Aghanim:2018eyx}. This sets a lower bound on $Z$ of order (neglecting DR from inflaton decay since this is diluted by the decay of $\phi$) \cite{Cicoli:2015bpq}:
\be
\Delta N_{\rm eff}\simeq  3\,\frac{\Gamma_{\phi\to{\rm DR}}}{\Gamma_{\phi\to{\rm vis}}}= \frac{3}{Z^2}\lesssim 0.75 \quad\text{for}\quad Z\gtrsim 2\,.
\ee
Therefore the ratio $\Gamma_\phi/\Gamma_{\phi\to{\rm vis}}$ which appears in (\ref{DMrelab1}) for the final DM abundance looks like:
\be
\frac{\Gamma_\phi}{\Gamma_{\phi\to{\rm vis}}} = 1+ \frac{1}{Z^2}\lesssim 1.25\,,
\ee
and the reheating temperature $T_{\rm R}$ in (\ref{TR}) can be derived from the following decay width:
\be
\Gamma_\phi = \frac{1+Z^2}{48\pi}\frac{m_\phi^3}{M_{\rm P}^2}\,.
\label{Moduluswidth}
\ee
Finally, the remaining quantities which are crucial to derive the fractional energy density of DR using (\ref{DRrelab1}) are:
\bea
\Gamma_\sigma &=& N_g^{\rm hid}\,\frac{c_{\rm hid}^2}{64\pi}\left(1+\frac{N_g}{N_g^{\rm hid}}\frac{1}{c_{\rm hid}^4}\right)\frac{m_\sigma^3}{M_{\rm P}^2} \nn \\
&\simeq& N_g^{\rm hid}\,\frac{c_{\rm hid}^2}{64\pi}\,\frac{m_\sigma^3}{M_{\rm P}^2}\,,
\label{InflatonWidth}
\eea
and:
\be
\frac{\Gamma_{\sigma\to{\rm DR}}}{\Gamma_\sigma} = \left(1+\frac{N_g}{N_g^{\rm hid}}\frac{1}{c_{\rm hid}^4}\right)^{-1}\simeq 1\,.
\ee

\subsection{Consistency of the branching scenario}

In this paper we are considering a branching scenario for DM production from inflaton decay. This is generically the case for superheavy WIMP DM since the corresponding annihilation rate would be too small to realize the so-called non-thermal annihilation scenario. However the computation of the DM relic density relies on the assumption that the standard thermal freeze-out mechanism cannot occur. This is true if the visible sector reheating temperature after the inflaton decay $T_{\rm R,inf}^{\rm vis}$ is below $T_{\rm f}$, where $T_{\rm f} \simeq m_{\chi}/20$ in the case of freeze-out (and $T_{\rm f} \simeq m_\chi/4$ for freeze-in). We shall now show that this is indeed the case in our model.

In standard supersymmetric scenarios, the LSP mass is expected to be of order the soft mass. As we have already seen, the DM mass is therefore slightly below the inflaton mass, $m_\chi \simeq m_\sigma/(\ln\epsilon)^2 < m_\sigma$. Notice that this feature is not a peculiarity of our model but it is a generic characteristic of string compactifications since, whenever the 4-cycle supporting the visible sector is stabilized in the geometric regime, the visible sector is always not sequestered from the source of supersymmetry breaking in the bulk. Thus the soft terms, and the DM mass, turn out to be of the same order as the gravitino mass, which sets also the order of magnitude of the mass of generic moduli (up to possible $|\ln\epsilon|$ suppression factors).

Therefore we shall consider $T_{\rm f} \simeq m_\sigma/[20 \,(\ln\epsilon)^2]$. On the other hand, the visible sector reheating temperature reads:
\bea
T_{\rm R,inf}^{\rm vis} &=& \left(\frac{40 N_g N_g^{\rm hid}}{\pi^2 g_*}\right)^{1/4} \sqrt{\frac{c_{\rm vis}c_{\rm hid}}{64\pi}} \,m_\sigma\, \sqrt{\frac{m_\sigma}{M_{\rm P}}} \nn \\
&\simeq& {m_\chi \over 20} \,(\ln\epsilon)^2 \sqrt{\frac{m_\sigma}{M_{\rm P}}}\,,
\eea
where we used $c_{\rm hid} \simeq c_{\rm vis}^{-1}$, and $N_g^{\rm hid} \simeq N_g = 12$ . Hence we obtain $T_{\rm R,inf}^{\rm vis} < T_{\rm f}$ provided that:
\be
\frac{T_{\rm R,inf}^{\rm vis}}{T_{\rm f}} \simeq
(\ln\epsilon)^2 \sqrt{\frac{m_\sigma}{M_{\rm P}}} \simeq \kappa^{1/4}\, |\ln\epsilon|^{5/2}\, \sqrt{\epsilon}< 1\, .
\ee
This is indeed the case for $\epsilon\ll 1$ and $\kappa\ll 1$, which guarantees the consistency of the branching scenario. This will be confirmed in Sec. \ref{numerical_evolution} which presents a numerical analysis of the cosmological evolution of our model.

\section{Cosmology of the string model}
\label{Sec5}

In this Section we shall first determine the values of the microscopic parameters which give the right amplitude of the density perturbations and the correct DM abundance, finding a DM mass around $10^{10}$-$10^{11}$ GeV. We shall then perform a numerical analysis of the cosmological evolution of our string model with an epoch of modulus domination. 

\subsection{Inflationary observables and DM abundance}

Let us derive the allowed DM mass range in a single modulus cosmology. We achieve a rather precise prediction by imposing a combination of observational and geometrical constraints. We start with the expression for the number of e-foldings between horizon exit and the end of inflation \cite{Cicoli:2016olq}:
\be
N_{\rm e} \simeq 57 + \frac{1}{4}\ln r -\frac{1}{4}N_{\text{reh}} -\frac{1}{4} N_{\phi} +\frac{1}{4}\ln\left(\frac{\rho_{\sigma,\text{start}}}{\rho_{\sigma,\text{end}}} \right).
\label{Nefold}
\ee
Here $r$ is the tensor-to-scalar ratio, $N_{\text{reh}}$ is the duration of the reheating period due to the inflaton $\sigma$, and $N_{\phi}$ is the duration of the EMD epoch due to the modulus $\phi$. Note that we have set the equation-of-state parameter $w$ equal to zero during inflationary reheating. Also $\rho_{\sigma,\text{start}}$ is the energy density at horizon exit, while $\rho_{\sigma,\text{end}}$ is the energy density at the end of inflation. Let us rewrite (\ref{Nefold}) in terms of fundamental parameters. The duration of the reheating period is:
\be
N_{\text{reh}} \simeq \frac{2}{3}\ln\left( \frac{H_{\sigma,\text{end}}}{\Gamma_{\sigma}} \right),
\ee
where $H_{\sigma,\text{end}}$ is the Hubble rate at the end of inflation, which is given by \cite{Cicoli:2016olq}:
\be
H_{\sigma,\text{end}} \simeq \sqrt{\frac{3}{2}\frac{\kappa}{(2\pi)^{3/2}\,W_0}} \, \epsilon^{3/2} |\ln\epsilon|^{3/4}\,M_{\rm P}\,.
\label{Hend}
\ee
Combining (\ref{Hend}) with the inflaton decay rate (\ref{InflatonWidth}) gives:
\be
N_{\text{reh}} \simeq \frac{2}{3}\ln\left(  \sqrt{\frac{3}{2}\frac{512^2\pi^4}{(2\pi)^{3/2}}} \frac{\mathcal{V}^{1/2}}{N_g^{\text{hid}}W_0^2g_s^{5/2}\left|\ln\epsilon\right|^{9/4}}\right).
\ee
The duration of modulus domination is given by:
\bea
N_{\phi} &\simeq& \frac{2}{3}\ln\left(Y_{\phi}^4\,\frac{\Gamma_{\sigma}}{\Gamma_{\phi}} \right) \\
&\simeq& \frac{2}{3}\ln \left(\frac34\,\frac{N_g^{\text{hid}}}{1+Z^2}\,Y_{\phi}^4\,g_s^{15/4}\,\vo^{5/2}\left|\ln\epsilon\right|^{9/2} \right). \nn
\eea
Following again \cite{Cicoli:2016olq}, the tensor-to-scalar ratio can be expressed as:
\be
r \simeq 16\times 3.7 \times 10^6 \left(\frac{3}{2}\frac{\left|\ln\epsilon\right|^{3/2}}{(2\pi)^{3/2}}\right)\frac{g_s  }{16\pi}\frac{W_0^2}{\mathcal{V}^3}\,.
\ee
Noting that the amplitude of the density perturbations can be written as $A_s=\frac{2}{3\pi^2 r}\frac{\rho_{\sigma,\text{start}}}{M_{\rm P}^4}$, we get:
\be
N_{\rm e} \simeq 60.1 - \frac{1}{6}\ln\left(\frac{Y_{\phi}^4\,\mathcal{V}^{15/2}}{5g_s^{1/4}W_0^5\left|\ln\epsilon\right|^{9/4}} \right),
\label{NNefold}
\ee
where we have set $Z=2$ and used $\ln\left(10^{10}A_s \right) = 3.044$ \cite{Planck}. To proceed further, we need the relation between the inflaton $\tau_{\rm inf}$, the volume $\vo$, and the number of e-foldings $N_{\rm e}$ that matches the observed value of $A_s$. This reads \cite{Cicoli:2016olq}:
\be
\tau_{\rm inf} \simeq 1.15\times 10^{-11}\frac{1}{ 2\pi g_s^2}\left(\frac{\mathcal{V}}{W_0\left|\ln\epsilon\right|^{3/4}N_{\rm e}}\right)^4.
\ee
Given that $\tau_{\rm inf}$ describes the volume of a local 4-cycle, we have to impose the geometrical constraint $\mathcal{V}^{2/3}\simeq \tau_b\gg \tau_{\rm inf}$ which guarantees that the effective field theory is under control. We can implement this constraint as $\vo^{2/3}\simeq \lambda\,\tau_{\rm inf}$, where $\lambda\gg 1$ is a tunable parameter that determines the hierarchy between the overall volume $\vo$ and the volume of the blow-up mode $\tau_{\rm inf}$. This gives us the final expression:
\be
\vo^{2/3} \simeq \lambda \left(\frac{\alpha^{1/4}\vo}{g_s^{1/2}W_0\left|\ln\epsilon\right|^{3/4}N_{\rm e}}\right)^4,
\label{boxed}
\ee
with $\alpha = 1.15\times 10^{-11}\frac{1}{ 2\pi }$ and $N_{\rm e}$ given in (\ref{NNefold}).

Let us briefly summarize the procedure that we shall follow to derive the DM mass corresponding to the observed DM abundance:
\begin{itemize}
\item We extract from (\ref{boxed}) $W_0$ as a function of $\vo$. This step encodes in $W_0(\vo)$ the information of the amplitude of the density perturbations and the geometrical relation between $\vo^{2/3}$ and $\tau_{\rm inf}$. We also set a natural bound on $W_0$ by constraining it to be in the range $\mathcal{O}(1-10^3)$.

\item We perform this step for different values of the underlying parameters $g_s$, $Y_\phi$, and $\lambda$, choosing the discrete parameter space to be $g_s \in [10^{-3},0.1]$, $Y_{\phi} \in [0.01,1]$, and $\lambda \in [10,10^4]$. We start with $1444$ initial combinations.

\item We extract the value of $\vo$ by matching the expressions for the observed and predicted DM abundances. We perform this step for each of the $1444$ initial parameter combinations.

\item We compute the DM mass for those parameter combinations that allow for the correct DM abundance.
\end{itemize}

\begin{figure}[ht!]
    \centering
     \includegraphics[trim=0cm 0cm 0cm 0cm, clip=true, width=0.48\textwidth]{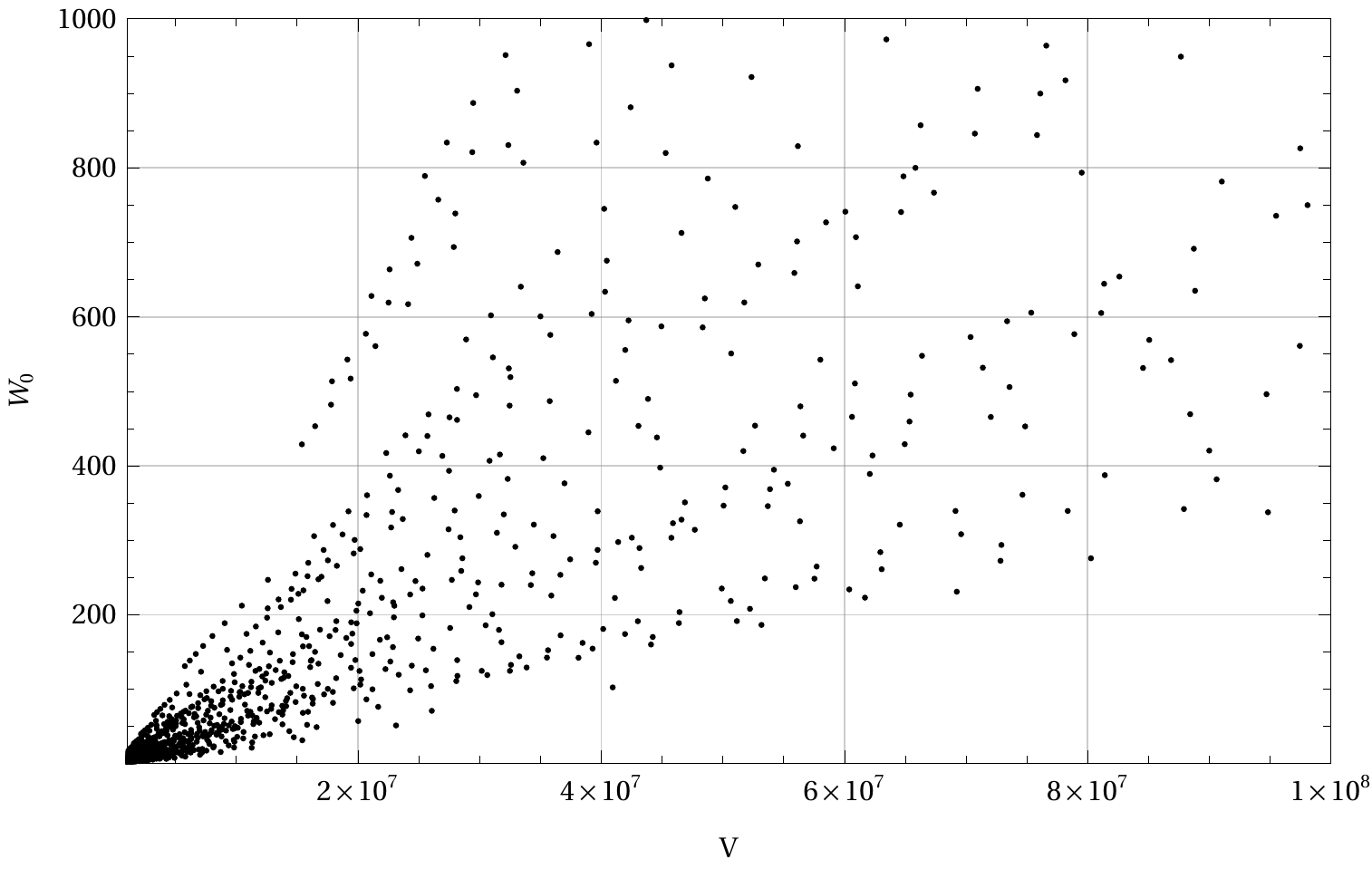}
    \caption{Points in the $(W_0,\vo)$ plane which reproduce the observed amplitude of the density perturbations and DM abundance.}
\label{Fig1}
\end{figure}

In Fig. \ref{Fig1} we present all data points in the $(W_0,\vo)$ plane which reproduce the observed amplitude of the density perturbations, respect our geometrical constraints, and yield the correct DM abundance. Approximately $72\%$ of our initial parameter space leads to a consistent solution. Notice that, although each point corresponds to different values of $g_s$, $Y_\phi$, and $\lambda$, the resulting DM mass is always in the range $10^{10}$-$10^{11}$ GeV, giving a robust prediction which is almost independent of the variation of the underlying parameters.

\begin{figure}[t]
\centering
     \includegraphics[trim=0cm 0cm 0cm 0cm, clip=true, width=0.48\textwidth]{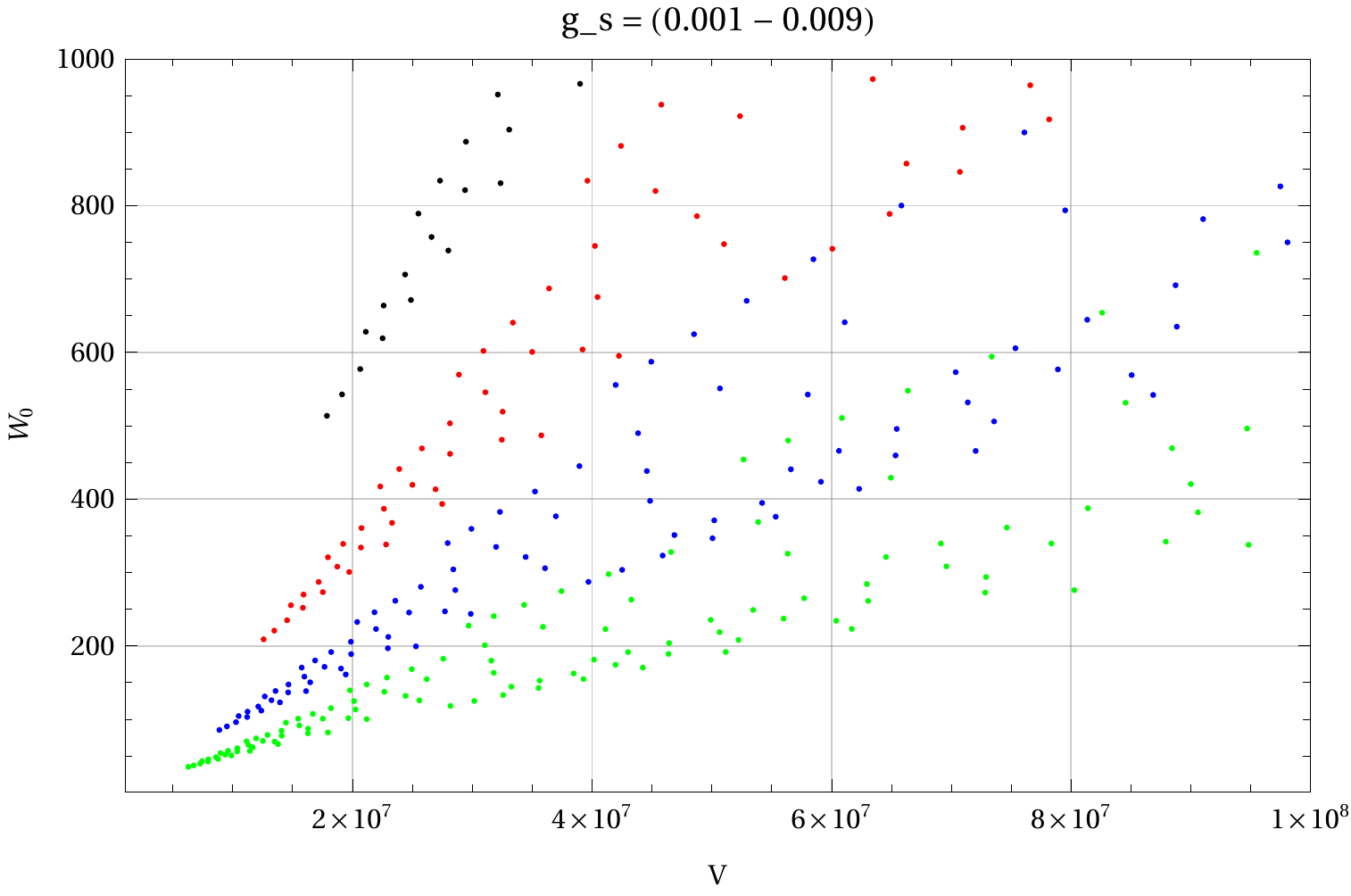}\\
      \includegraphics[trim=0cm 0cm 0cm 0cm, clip=true, width=0.48\textwidth]{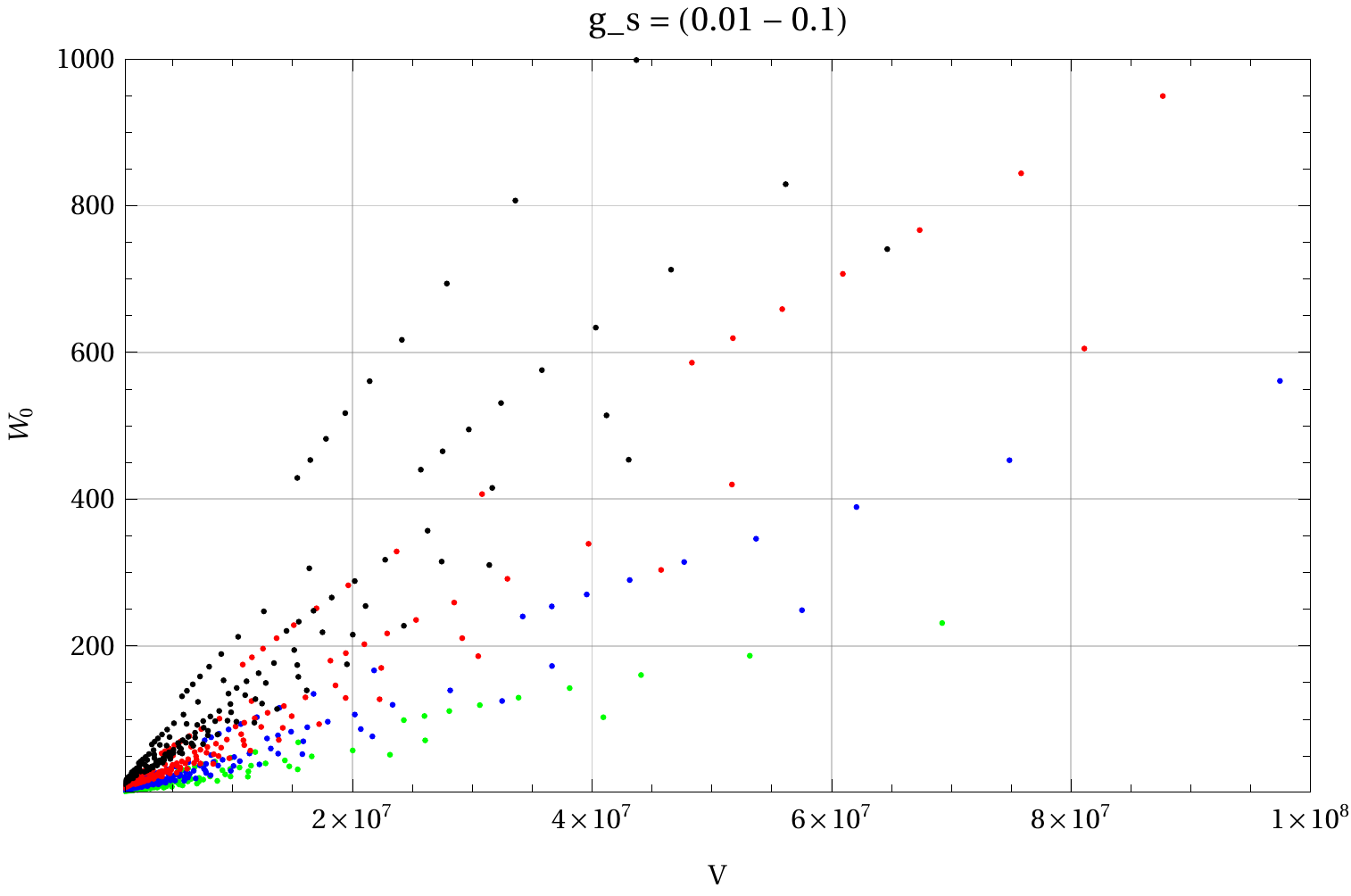}
      \caption{Selected data points for different values of $g_s$ and $\lambda$. Around 29\% of the data points are in the lower $g_s$ regime, while around 71\% are in the upper regime. Black points correspond to $\lambda=10^4$, red to $\lambda=10^3$, blue to $\lambda=10^2$, and green to $\lambda=10$. 
      \vspace{-0.1cm}}    
\label{Fig2}
\end{figure}

The accumulation around the origin and the jet-like structures in the distribution of the data points can be understood from Fig. \ref{Fig2} where we split the points shown in Fig. \ref{Fig1} into two sets with, respectively, $g_s=0.001-0.009$ and $g_s=0.01-0.1$. Moreover black dots correspond to $\lambda=10^4$, red to $\lambda=10^3$, blue to $\lambda=10^2$, and green to $\lambda=10$. The plot for smaller values of $g_s$ shows clearly that the four jet structures correspond to different values of $\lambda$. This behavior is a direct consequence of (\ref{boxed}) which implies that $\lambda$ determines the slope of the function $W_0(\vo)$. On the other hand, the plot for larger values of $g_s$ features a larger density at smaller values of $W_0$ and $\vo$. This behavior is a consequence of the consistency of the branching scenario. In fact, in order for (\ref{ratiocvischid}) to hold, smaller values of the volume must lead to an increase in $g_s$. It is worth mentioning also that around 71\% of the data points correspond to $g_s \in [0.01,0.1]$, whereas 29\% of the acceptable parameter space correspond to $g_s \in [10^{-3},0.01]$. 

\begin{figure}[ht!]
\centering
     \includegraphics[trim=0cm 0cm 0cm 0cm, clip=true, width=0.48\textwidth]{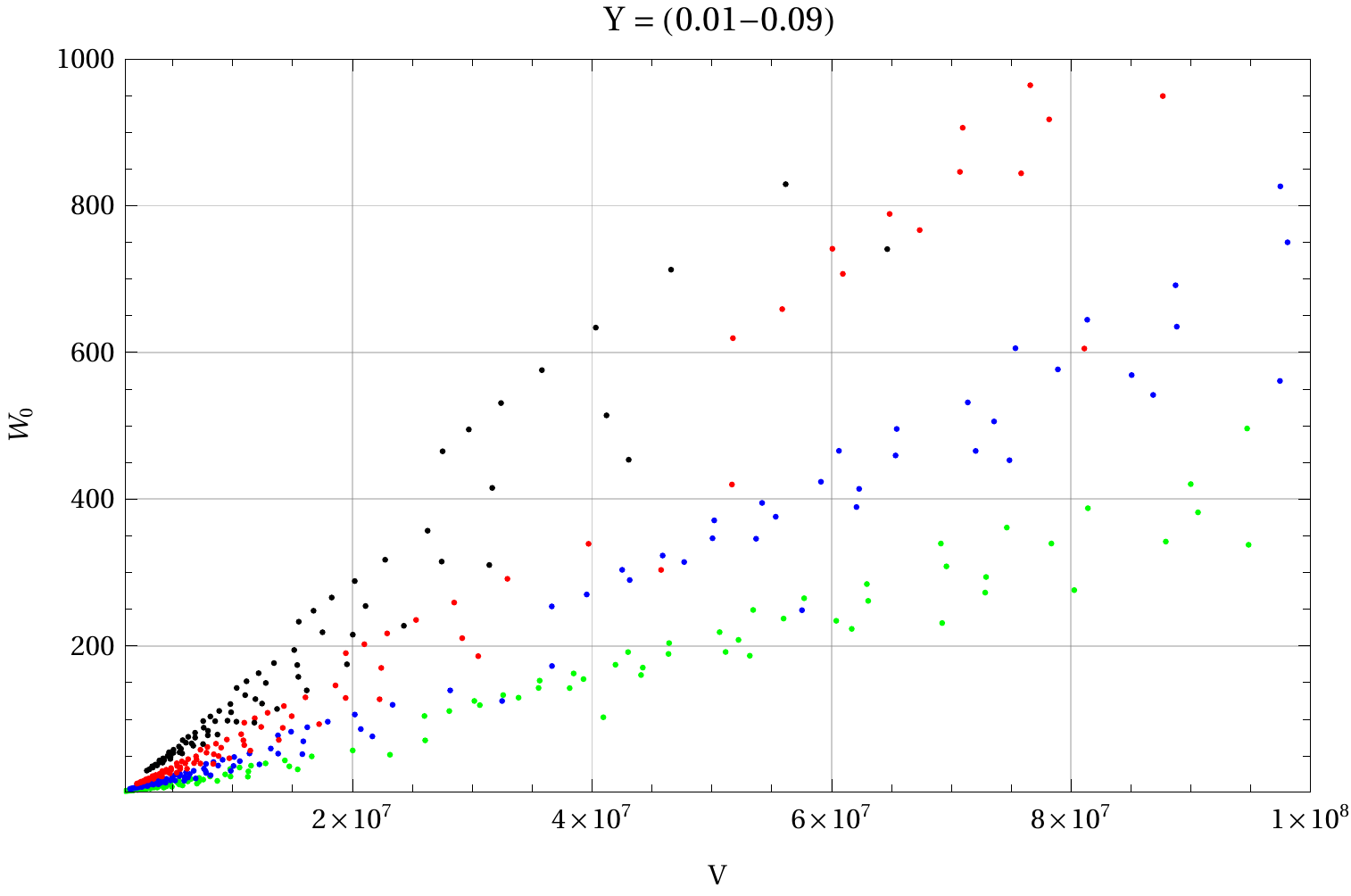}\\
      \includegraphics[trim=0cm 0cm 0cm 0cm, clip=true, width=0.48\textwidth]{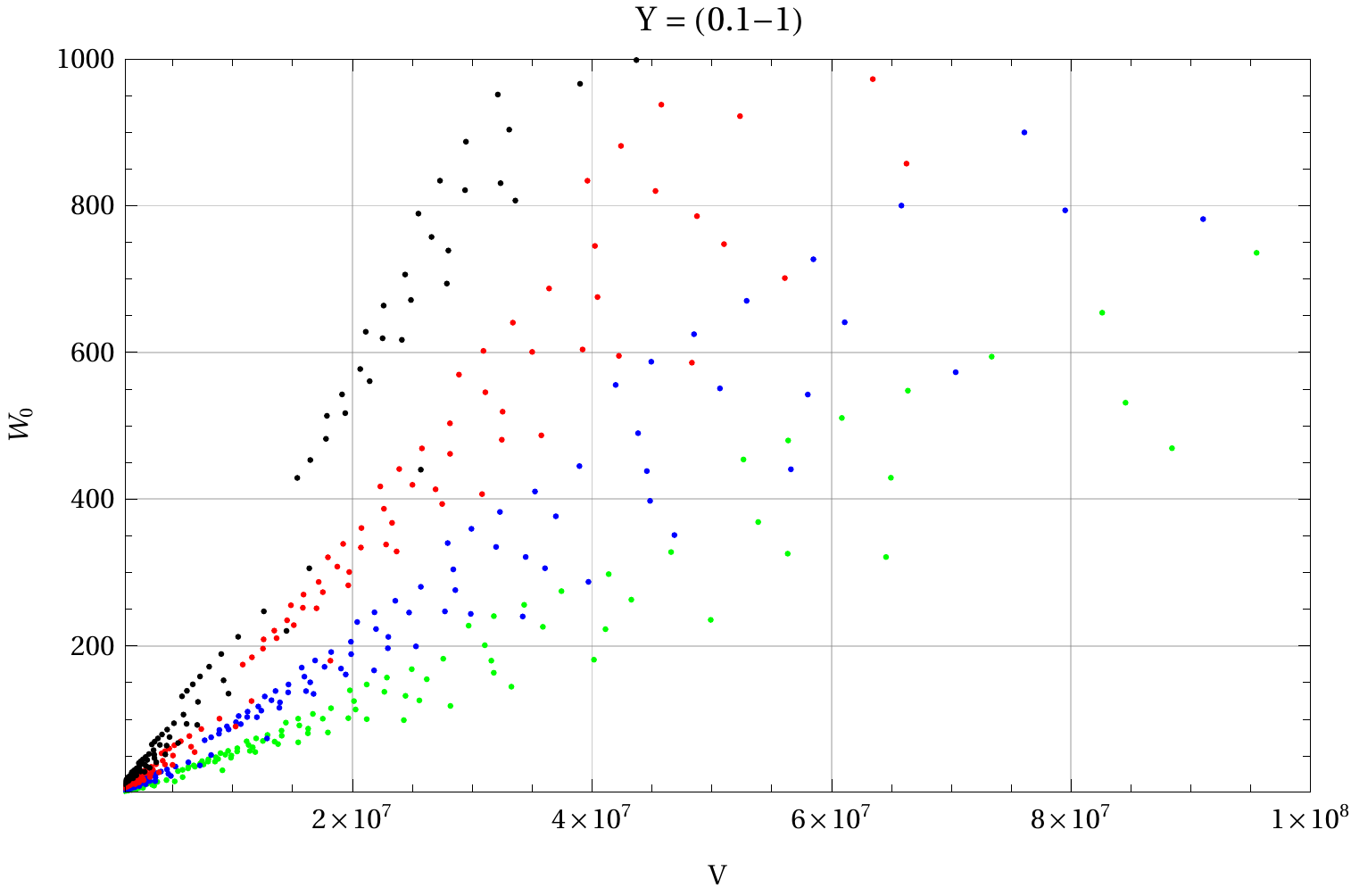}   
      \caption{Selected data points for different values of $Y_\phi$ and $\lambda$. Around 42\% of the data points are in the lower $Y_\phi$ regime, while around 58\% are in the upper regime. Black points correspond to $\lambda=10^4$, red to $\lambda=10^3$, blue to $\lambda=10^2$, and green to $\lambda=10$. }    
\label{Fig3}
\end{figure}

In Fig. \ref{Fig3} we present a similar analysis, this time splitting all data points from Fig. \ref{Fig1} into two sets, depending on the value of the misalignment $Y_\phi$. We observe again the same jet structure depending on the value of the parameter $\lambda$. An important observation here is the slight rotation of the data-point cone towards the $W_0$ axis if we increase $Y_\phi$. This behavior is mainly driven by the DM abundance constraint formulated in (\ref{DMrelab1}). The abundance scales like $\sim {Y_\phi}^{-2}\vo^{-13/4}$. Hence, in order to match the right abundance, a smaller volume must be compensated by a larger misalignment $Y_\phi$.

\begin{figure}[ht!]
\centering
     \includegraphics[trim=0cm 0cm 0cm 0cm, clip=true, width=0.48\textwidth]{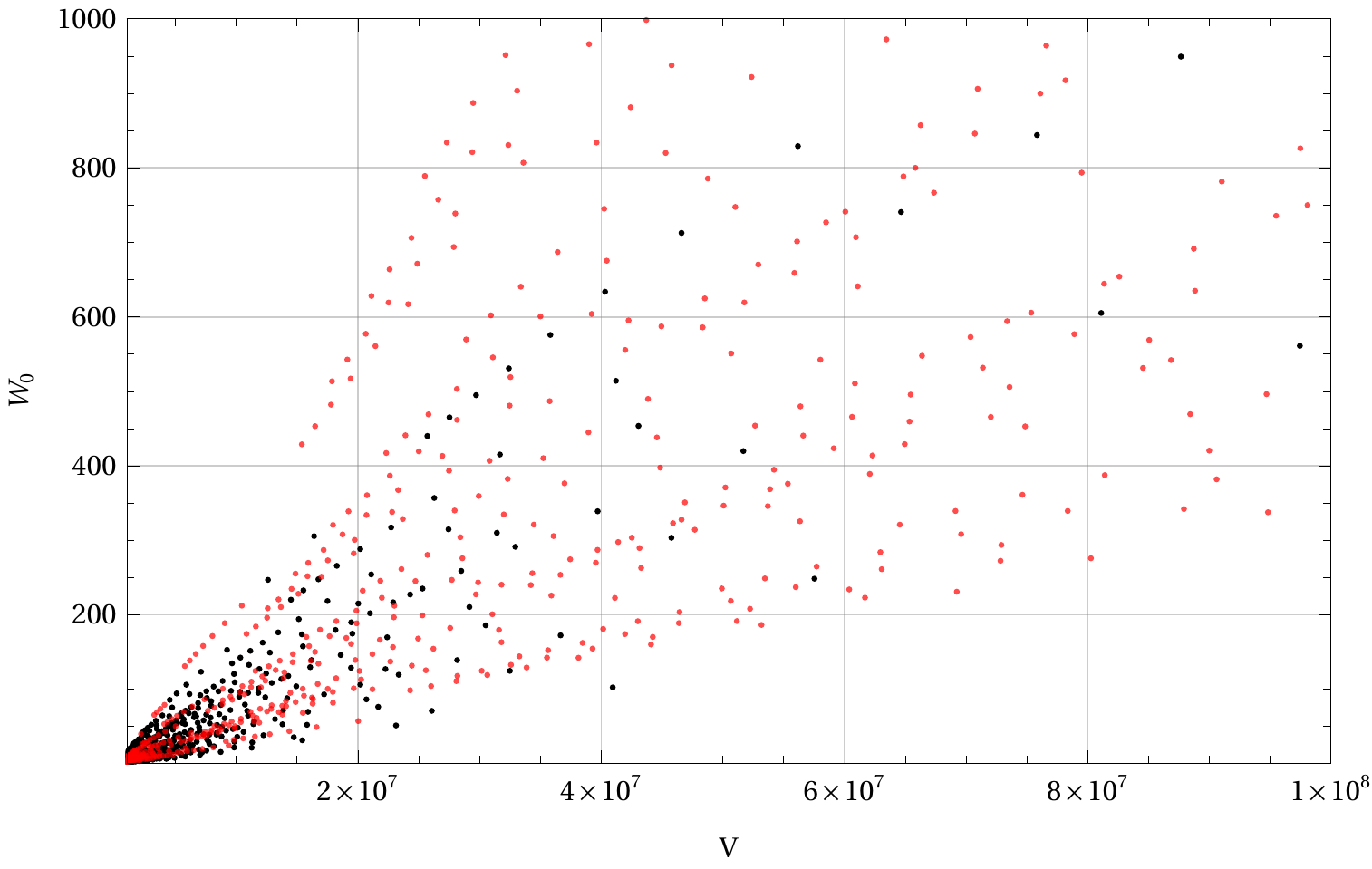}  
      \caption{Scalar spectral index coloring of the entire data set. Black dots correspond to $n_{\rm s}$ within the $2\sigma$ range, while red dots have $n_{\rm s}$ within $3\sigma$ (but outside $2\sigma$).
      \vspace{-0.3cm}}    
\label{Fig4}
\end{figure}

Understanding the behavior of our data set as a function of the underlying parameters is important in order to understand the distribution of the scalar spectral index $n_{\rm s}$. For each point in Fig. \ref{Fig1}, we calculated the resulting value of $n_{\rm s}$. All obtained values are within the 2- and 3$\sigma$ range \cite{Planck}, as can be seen from Fig. \ref{Fig4}.

In Fig. \ref{Fig4} each black dot corresponds to a scalar spectral index within the $2\sigma$ range, i.e. $ 0.9565 <n_{\rm s}< 0.9733$, while each red dot has $n_{\rm s}$ in the $3\sigma$ range, i.e. $ 0.9523 <n_{\rm s}< 0.9775$, but outside $2\sigma$. By comparing the two distributions, we see that the data points corresponding to $2\sigma$ seem to be accumulating at the origin and their cone is slightly rotated towards the volume axis. From this observation we conclude that phenomenologically more acceptable values of $n_{\rm s}$ drive the string coupling $g_s$ to larger values and the misalignment $Y_\phi$ to smaller ones. 

A crucial observation is that the parameter values in our data set naturally respect the relation between the volume and the underlying model parameters at the minimum of the scalar potential \cite{LVS}:
\be
\langle \vo \rangle \simeq \frac{3\sqrt{\langle \tau_{\rm inf} \rangle}|W_0|}{4a_{\rm inf}A_{\rm inf}}e^{a_{\rm inf}\langle \tau_{\rm inf} \rangle}, \quad \langle \tau_{\rm inf} \rangle \simeq \frac{1}{g_s}\left(\frac{\xi}{2} \right)^{2/3}, \nn
\ee
for natural $\mathcal{O}(1)$ values of the microscopic parameters $a_{\rm inf}$, $A_{\rm inf}$ and $\xi$.

In Sec. \ref{numerical_evolution} we shall perform a more in-depth numerical analysis of the cosmological evolution, using the benchmark parameters listed in Tab. \ref{Tab1}.

\begin{table}[ht!]
 \begin{tabular}{||c || c | c ||}
\hline
$W_0$ & 39.1  \\
\hline
$\vo$ & $8.4 \times 10^6$ \\
 \hline
$N_e$ & 47.4\\
\hline
$N_{\text{reh}}$ & 3.7 \\
\hline
$N_{\phi}$ & 16.4\\
\hline
$n_{\rm s}$ & 0.9578\\
\hline
$m_{\sigma}$ & $8.7 \times 10^{12} \text{GeV}$ \\
\hline
$m_{\phi}$ & $3.9 \times 10^8\text{GeV}$\\
\hline
$m_{3/2}$ & $7.1 \times 10^{11}\text{GeV}$\\
\hline
$m_{\chi}$ & $5.8 \times 10^{10} \text{GeV}$\\
\hline
$c_{\text{hid}}$ & 514.7\\
\hline
\end{tabular}
\caption{Microscopic parameters $W_0$ and $\vo$, the resulting e-folding numbers, $n_{\rm s}$, mass scales, and inflaton coupling to hidden degrees of freedom $c_{\rm hid}$ at a benchmark point that gives the right amplitude of the density perturbations and the correct DM abundance. The input parameters are $g_s = 0.1$, $Y_{\phi} = 0.01$, $\lambda = 10^3$, $N_g^{\rm hid}=12$, and $Z = 2$.}
\label{Tab1}
\end{table}

\vspace{-0.7cm}
\subsection{Numerical analysis of cosmological evolution}
\label{numerical_evolution}

We perform a numerical analysis of the cosmological evolution of our scenario by solving the coupled set of Boltzmann equations for the various cosmological components (see App. \ref{App} for a scenario with two moduli). We begin the numerical evolution at \(H \simeq H_{\rm inf}\), with both \(\sigma\) and \(\phi\) oscillating, and other components highly subdominant. The Boltzmann equations for our single-modulus scenario are as follows:
\begin{align}
& \frac{d\rho_\sigma}{dt} + 3H\rho_\sigma = - \Gamma_\sigma\,\rho_\sigma \, ,
    \label{boltz1_inf}
\\
& \frac{d\rho_{\phi}}{dt} + 3H\rho_{\phi} = - \Gamma_{\phi}\, \rho_{\phi} \, ,
    \label{boltz1_phi}
\\
& \frac{d\rho_{\rm DR}}{dt} + 4H\rho_{\rm DR} = \Gamma_{\rm\sigma \rightarrow DR}\rho_\sigma + \Gamma_{\rm\phi \rightarrow DR}\rho_{\phi} \, ,
    \label{boltz1_DR}
\\
& \frac{d\rho_{\rm R}}{dt} + 4H\rho_{\rm R} = \Gamma_{\rm \sigma \rightarrow vis}\rho_\sigma + \Gamma_{\phi \to {\rm vis}}\rho_{\phi} \, ,
    \label{boltz1_R}
\\
& \frac{dn_\chi}{dt} + 3Hn_\chi = {\rm Br}_\chi\Gamma_\sigma \left(\frac{\rho_\sigma}{m_\sigma}\right) + \left<\sigma_{\rm ann}v\right>\left(n_{\rm\chi,eq}^2 - n_\chi^2\right), 
\label{boltz1_chi}
\end{align}
where the Hubble rate \(H\) is given by the sum of all energy density components, and the various decay rates are given in \eqref{ratiocvischid}, \eqref{Moduluswidth} and \eqref{InflatonWidth} using the benchmark values of Tab. \ref{Tab1}. $\langle \sigma_{\rm ann} v \rangle$ denotes the thermally averaged rate for $\chi$ production from/annihilation to the thermal bath with the average energy per \(\chi\) particle approximated as \(\left<E_\chi\right> \approx \sqrt{m_\chi^2 + 9T_{\rm vis}^2}\) \cite{GKR:2000ex}. Here, we take \(\left<\sigma_{\rm ann}v\right> \approx \alpha_\chi^2/m_\chi^2\) with \(\alpha_\chi \sim 0.1\). This happens to be the case, for example, for Higgsino and Wino DM~\cite{split}. However, because thermal production is subdominant in our scenarios, the exact form of \(\left<\sigma_{\rm ann}v\right>\), including possible temperature dependence, is not really important. For typical DM masses in our scenarios, $m_\chi \sim 10^{10}$-$10^{11}$ GeV, we obtain values of $\langle \sigma_{\rm ann} v \rangle$ in the freeze-in regime. Finally, the DM equilibrium number density, relevant for thermal production, is given by:
\be
n_{\rm\chi,eq} = \frac{g_\chi}{\left(2\pi\right)^3}\int\frac{d^3p}{e^{E(p)/T_{\rm vis}} \pm 1}\,.
\ee

\begin{figure}[ht]
    \centering
    \includegraphics[trim=0cm 0cm 0cm 0cm, clip=true, width=0.48\textwidth]{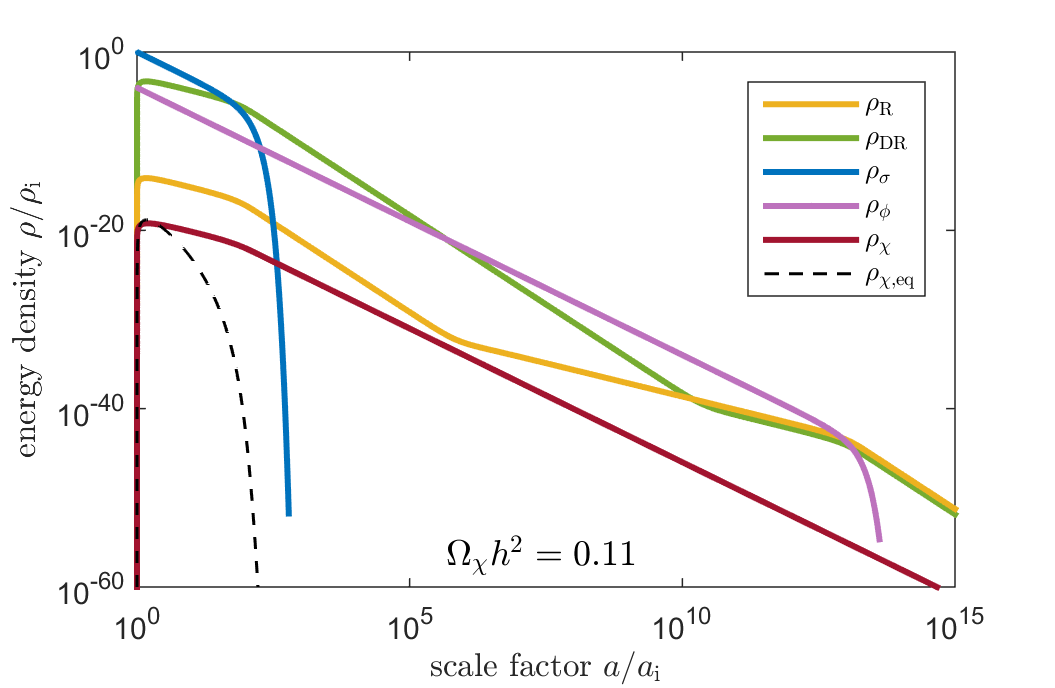}
    \caption{Numerical evolution of the system in \eqref{boltz1_inf}-\eqref{boltz1_chi}. Curves depict the energy densities as functions of scale factor in our scenario. Numerical values of the underlying parameters correspond to the benchmark values given in Tab. \ref{Tab1}. DM is primarily produced from inflaton decay, with a negligible thermal contribution, establishing the observed relic abundance. }
    \label{fig:1mod_rho}
\end{figure}

A sample numerical solution of \eqref{boltz1_inf}-\eqref{boltz1_chi} is shown in Fig. \ref{fig:1mod_rho}. As the evolution proceeds, DM, dark radiation, and ordinary radiation are continually produced by inflaton decay until \(H \simeq \Gamma_\sigma\), at which point inflaton decay completes. This begins an era of hidden-radiation domination which lasts until the light modulus \(\phi\) overcomes the energy density of hidden radiation. From here until the time when \(H \simeq \Gamma_\phi \), we have a period of EMD driven by the modulus, which is then followed by the standard period of radiation domination once the modulus decay completes.

\begin{figure}[htb]
    \centering
    \includegraphics[trim=0cm 0cm 0cm 0cm, clip=true, width=0.48\textwidth]{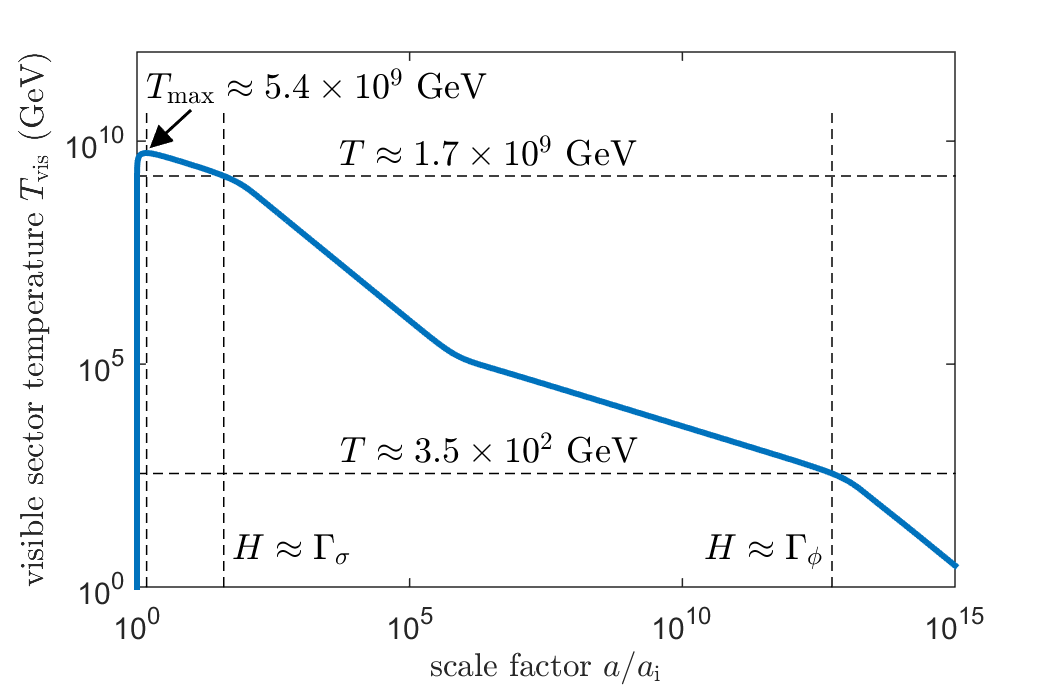}
    \caption{Evolution of the visible sector temperature as a function of scale factor in the cosmological history of Fig. \ref{fig:1mod_rho}.
    \vspace{-0.1cm}}
    \label{fig:1mod_HT}
\end{figure}

The DM abundance is set by the inflaton decay at \(H \simeq \Gamma_\sigma\) and simply redshifts through the remaining cosmological history. For typical values of the parameters in our scenario, the maximum visible sector temperature established during inflationary reheating is smaller than the DM mass, such that thermal production of DM occurs on the Boltzmann tail of the equilibrium distribution, rendering the thermal contribution irrelevant.\footnote{Freeze-in production of DM from the visible sector thermal bath is quite sensitive to the DM mass, and can dominate over the branching contribution from inflaton decay if the DM mass is lowered below the range in our scenario.} Fig. \ref{fig:1mod_HT} shows the visible sector temperature as a function of the scale factor for the cosmological history shown in Fig. \ref{fig:1mod_rho}, where we have assumed a smooth function for the temperature dependence of the relativistic degrees of freedom in the visible sector.

One comment is in order at this point. Our calculation of freeze-in production of DM in (\ref{boltz1_chi}) assumes instantaneous thermalization of inflaton decay products in the visible sector. In fact, it holds as long as the visible sector reaches thermal equilibrium at a temperature $T > T_{\rm f}$. However, due to the small number density of inflaton decay products in the visible sector, thermalization may be significantly delayed (for example, see~\cite{thermal1,thermal2}). If $T < T_{\rm f}$ at the time of thermalization, then thermal production of DM will be completely negligible. Before thermal equilibrium is established, DM production from inflaton decay products is kinematically possible due to their typical mass hierarchy $m_\sigma \gg m_\chi$~\cite{thermal3,thermal4,thermal5,thermal6}. However, by conservation of energy, the number density of these particles is much smaller than what it would be in thermal equilibrium. We have checked that DM production during thermalization is a few orders of magnitude smaller than that from direct inflaton decay, for the parameters shown in Tab. \ref{Tab1}, even if the visible sector is not thermalized until $H \simeq \Gamma_\sigma$.       

\section{Conclusions}
\label{Sec6}

In this paper we have argued that two generic features of string compactifications, a high supersymmetry breaking scale (which is favored by both statistical arguments \cite{Denef:2004ze, Broeckel:2020fdz} and by the requirement of a viable inflationary model building \cite{Kallosh:2004yh}) and the presence of light moduli which drive epochs of EMD \cite{KSW, Bobby, Cicoli:2016olq}, lead typically to superheavy WIMP DM with mass around the intermediate scale. This scenario has not received significant attention so far because DM with a mass in the $10^{10}$-$10^{11}$ GeV range is inevitably overproduced in a standard thermal history. However, if DM is produced non-thermally from the decay of the inflaton and it is subsequently diluted by the decay of long-lived string moduli which are so light that their decay does not reproduce DM, one can obtain the observed abundance in this so-called branching scenario even for very large DM masses. This does not just account for the non-observation of supersymmetry and WIMPs at colliders, but it may also provide a natural explanation of the origin of ultra-high-energy cosmic rays recently observed by IceCube and ANITA, if DM is unstable and has the right coupling to neutrinos \cite{Heurtier:2019git}. 

We illustrated this general picture by presenting two explicit 4D string models which lead to superheavy WIMP DM. The first model is described in Sec. \ref{Sec4} and features a single epoch of modulus domination, while App. \ref{App} gives all the details of a different model with two epochs of EMD driven by two different light moduli. It turns out that in both cases the observed DM abundance can be obtained for a mass around $10^{10}$-$10^{11}$ GeV. The main virtue of both models is the possibility to follow their entire cosmological evolution from inflation to the final reheating (due to the decay of the lightest modulus) that establishes a RD universe before the onset of BBN. This can be achieved by focusing on type IIB LVS string models where the exponentially large volume of the extra dimensions allows to keep control over the 4D low-energy effective field theory. Hence all moduli masses and couplings to both visible and hidden sector degrees of freedom can be computed in detail. Moreover one can build 4D models which can realize inflation, supersymmetry breaking and a chiral MSSM-like visible sector on D-branes (see \cite{Cicoli:2011qg, Cicoli:2012vw, Cicoli:2013mpa, Cicoli:2013cha, Cicoli:2017axo, Cicoli:2017shd} for explicit Calabi-Yau models with all these features). 

We followed the entire cosmological evolution in both models using analytical and numerical tools. This allowed us to combine various constraints coming from both theoretical and phenomenological considerations. Interestingly, we derived the ranges of the microscopic parameters in a regime where geometrical constraints on the underlying extra-dimensional construction are respected, which yield the observed DM abundance as well as the correct value of inflationary observables, namely the amplitude of the density perturbations and the scalar spectral index. 

Future investigations could include more formal aspects as well as more phenomenological implications of our findings. From the formal point of view, it would be very interesting to investigate how generic superheavy WIMP DM is from the string landscape point of view, for example comparing this scenario to the case of fuzzy DM \cite{Hui:2016ltb} which has also been claimed to be a natural outcome of string models due to the ubiquitous presence of ultra-light axions \cite{Svrcek:2006yi, Nflation, Arvanitaki:2009fg, Cicoli:2012sz}. On the other hand, from the phenomenological side, it is crucial to understand how a superheavy DM could be detected in actual observations, for example establishing in more detail the possible connection of our results with the production of very energetic cosmic rays from DM decay. We leave all these intriguing possibilities for future work.

\section{Acknowledgements}

We would like to thank V. Guidetti and F. Muia for helpful discussions. The work of R.A. and J.O. is supported in part by NSF Grant No.  PHY-1720174.

\appendix

\section{Superheavy DM in a scenario with two moduli}
\label{App}

String compactifications are in general characterized by a large number of moduli and by a leading order no-scale structure that makes some of them (in general more than one) lighter than expected, i.e. $m_\phi\ll m_{3/2}$ \cite{Burgess:2020qsc}. Hence one might expect to have several epochs of modulus domination in the post-inflationary history. One could have either multiple eras of EMD separated by intermediate phases of RD (if there is a hierarchy among the initial displacements of the moduli), or a single extended EMD epoch (if the initial displacements of all moduli are of the same order). In what follows we shall investigate the features of this general scenario by focusing on the illustrative example with two light moduli.

\subsection{Branching scenario with two epochs of EMD}

Let us present a branching scenario that involves two moduli $\phi_1$ and $\phi_2$. The two moduli have masses $m_{\phi_1}$ and $m_{\phi_2}$ with $m_{\phi_2} \lesssim m_{\phi_1}$. They mainly decay to the visible sector with respective couplings $c_1/M_{\rm P}$ and $c_2/M_{\rm P}$, leading to decay widths $\Gamma_{\phi_1\to{\rm vis}} \simeq c^2_1 m^3_{\phi_1}/M^2_{\rm P}$ and $\Gamma_{\phi_2\to{\rm vis}} \simeq c^2_2 m^3_{\phi_2}/M^2_{\rm P}$ (we assume that their decays to hidden sector particles are suppressed and produce just a small amount of DR). Assuming that $m_{\phi_1} < m_\chi$, the decay of $\phi_1$ and $\phi_2$ to DM will be kinematically forbidden. Therefore their decay only dilutes the abundance of DM and DR produced from the inflaton decay. The initial displacements of $\phi_1$ and $\phi_2$ from the minimum of their potential are $\phi_{1,0}$ and $\phi_{2,0}$ respectively. We assume $\phi_{1,0} \gtrsim \phi_{2,0}$ so that each modulus can dominate the energy density of the universe for a period of time.

The important stages of the post-inflationary history in this scenario (in chronological order) are as follows:
\vskip 1.5mm
\noindent
{\bf 1-} $\Gamma_\sigma \lesssim H < H_{\rm inf}$: The universe is in an EMD phase driven by inflaton oscillations about the minimum of its potential.
Both moduli also start oscillating at this stage with respective energy densities $\rho_{\phi_1} = (\phi_{1,0}/M_{\rm P})^2 \rho_\sigma$ and $\rho_{\phi_2} = (\phi_{2,0}/M_{\rm P})^2 \rho_\sigma$. The inflaton decay completes at $H \simeq \Gamma_\sigma$ mainly populating the hidden sector.
\vskip 1.5mm
\noindent
{\bf 2-} $H_{{\rm D},1} \lesssim H < \Gamma_\sigma$: The universe is in a RD phase at this stage.
Moduli oscillations behave like matter, and hence $\rho_{\phi_{1,2}}$ redshifted more slowly than $\rho_{\rm R}$.
Since $\rho_{\phi_1} > \rho_{\phi_2}$, $\phi_1$ starts to dominate at $H_{{\rm D},1} \simeq (\phi_{1,0}/M_{\rm P})^4 \Gamma_\sigma$, which is the onset of a second phase of EMD.
\vskip 1.5mm
\noindent
{\bf 3-} $\Gamma_{\phi_1} \lesssim H < H_{{\rm D},1}$: The universe is in an EMD epoch that is driven by $\phi_1$ during this stage. Decay of $\phi_1$ completes when the Hubble expansion rate is $H \simeq \Gamma_{\phi_1}$ ($\Gamma_{\phi_1}$ denotes the total decay rate of $\phi_1$) and reheats the visible sector. This leads to the formation of a RD universe.
\vskip 1.5mm
\noindent
{\bf 4-} $H_{{\rm D},2} \lesssim H < \Gamma_{\phi_1}$: The universe is in an intermediate RD phase during this stage. Since $\rho_{\phi_2}$ is redshifted more slowly than $\rho_{\rm R}$, $\phi_2$ starts to dominate when the Hubble expansion rate is  $H_{{\rm D},2} \simeq (\phi_{2,0}/\phi_{1,0})^4 \Gamma_{\phi_1}$, which is the onset of another epoch of EMD.
\vskip 1.5mm
\noindent
{\bf 5-} $\Gamma_{\phi_2} \lesssim H < H_{{\rm D},2}$: The universe is in a third phase of EMD that is driven by $\phi_2$. The decay of $\phi_2$ completes  when the Hubble expansion rate is $H \simeq \Gamma_{\phi_2}$ ($\Gamma_{\phi_2}$ is the total decay rate of $\phi_2$), at which time the universe enters the final RD phase prior to the onset of BBN.
\vskip 1.5mm

One point to note is that $H_{{\rm D},2} \simeq \Gamma_{\phi_1}$ if $\phi_{2,0} \simeq \phi_{1,0}$. In this case, $\phi_2$ dominates the energy density of the universe as soon as the decay of $\phi_1$ completes. Stage 4 above thus effectively disappears and there is a direct transition from the first EMD era (stage 3) to the second one (stage 5), implying a single extended epoch of EMD driven by the two moduli $\phi_1$ and $\phi_2$.\footnote{This is an example of the two-field EMD scenario studied in~\cite{Jaksa1}.}

Let us now estimate the relic abundance of DM in this scenario. The number density of DM particles at $H \simeq \Gamma_{\phi_2}$ is given by:
\be
n_{\chi} \simeq n_\sigma ~ {\rm Br}_{\chi}  ~ \left({a_\sigma \over a_{{\rm D},1}}\right)^3  \left({a_{{\rm D},1} \over a_{\phi_1}}\right)^3  \left({a_{\phi_1} \over a_{{\rm D},2}}\right)^3  \left({a_{{\rm D},2} \over a_{\phi_2}}\right)^3 \, .
\label{nDM2}
\ee
This is similar to (\ref{nDM1}) for the case with a single epoch of modulus domination. The last two terms on the RHS, which are new, account for the dilution of the number density in stages 4 and 5 above respectively. After using the scaling of $a$ with time in stages 4 and 5 above, and normalizing $n_{\chi}$ by the entropy density $s$ at $H \simeq \Gamma_{\phi_2}$, we find:
\be
\frac{n_\chi}{s} \simeq \frac34 \times 10^{-3} ~ \frac{1}{Y_{\phi_2}^2}
~ \frac{\Gamma_{\sigma\to{\rm vis}}}{\Gamma_\sigma}
~ \frac{\Gamma_{\phi_2}}{\Gamma_{\phi_2\to{\rm vis}}}
~ \frac{T_{{\rm R},2}}{m_\sigma} \,,
\label{DMrelab2}
\ee
where:
\be
T_{{\rm R},2} = \left(\frac{90}{\pi^2 g_{*,{\rm R},2}}\,\frac{\Gamma_{\phi_2\to{\rm vis}}}{\Gamma_{\phi_2}}\right)^{1/4} \sqrt{\Gamma_{\phi_2} M_{\rm P}}\,, 
\label{TR2}
\ee
with $g_{*,{\rm R},2}$ denoting the number of relativistic degrees of freedom in the visible sector at $T = T_{{\rm R},2}$, and $Y_{\phi_2} \equiv \phi_{2,0}/M_{\rm P}$. The energy density of DR at $H \simeq \Gamma_{\phi_2}$ is given by (assuming that no DR is produced from $\phi_1$ decay):
\bea
\rho_{\rm DR} &\simeq& \rho_\sigma\,\frac{\Gamma_{\sigma\to{\rm DR}}}{\Gamma_\sigma} \left({a_\sigma \over a_{{\rm D},1}}\right)^4 \left({a_{{\rm D},1} \over a_{\phi_1}}\right)^4 \left({a_{\phi_1} \over a_{{\rm D},2}}\right)^4 \left({a_{{\rm D},2} \over a_{\phi_2}}\right)^4 \nn \\ 
&+& \rho_{\phi_2}\,\frac{\Gamma_{\phi_2\to{\rm DR}}}{\Gamma_{\phi_2}}.
\label{DR2}
\eea
This is similar to (\ref{DR1}) with two additional terms on the RHS that account for the energy density redshift in stages 4 and 5 respectively. Thus the fractional energy density of DR is given by:
\be
\label{DRrelab2}
{\rho_{\rm DR} \over \rho_{\rm R}}
\simeq \frac{1}{Y_{\phi_2}^{8/3}} \left(\frac{\Gamma_{\phi_2}}{\Gamma_\sigma}\right)^{2/3}\,\frac{\Gamma_{\sigma\to{\rm DR}}}{\Gamma_\sigma}\,\frac{\Gamma_{\phi_2}}{\Gamma_{\phi_2\to{\rm vis}}}+\frac{\Gamma_{\phi_2\to{\rm DR}}}{\Gamma_{\phi_2\to{\rm vis}}}\,.
\ee
Some comments are in order at this point. It is seen from (\ref{DMrelab2}) and (\ref{DRrelab2}) that final abundance of DM and DR in the two modulus scenario depends only on the initial amplitude and decay width of the second modulus $\phi_2$. This can be understood as follows. For fixed $Y_{\phi_2}$, varying $Y_{\phi_1} \equiv \phi_{1,0}/M_{\rm P}$ (as long as $Y_{\phi_1} \gtrsim Y_{\phi_2}$) affects the two epochs of modulus domination in opposite ways. Increasing (decreasing) $Y_{\phi_1}$ makes stage 4 longer (shorter) and stage 3 shorter (longer) by the same factor. A similar thing happens by decreasing (increasing) $\Gamma_{\phi_1}$ with $\Gamma_{\phi_2}$ fixed (as long as $\Gamma_{\phi_1} \gtrsim \Gamma_{\phi_2}$). As a result, the combined dilution factor from two epochs of modulus domination does not depend on the parameters of $\phi_1$.

That said, it is helpful to compare the DM and DR abundance with the previous scenario when there is one epoch of modulus domination. We can rewrite (\ref{DMrelab2}) in terms of (\ref{DMrelab1}) as follows (in the limit where the production of DR from the decay of the lightest modulus is completely negligible):
\be
\label{DMc}
\left.\frac{n_{\chi}}{s}\right|_2 = \left.\frac{n_{\chi}}{s}\right|_1 ~
\left({Y_{\phi_1} \over Y_{\phi_2}}\right)^2~ \left({T_{{\rm R},2} \over T_{{\rm R},1}}\right)\,.
\ee
Similarly, (\ref{DRrelab2}) can be written in terms of (\ref{DRrelab1}):
\be
\label{DRc}
\left.\frac{\rho_{\rm DR}}{\rho_{\rm R}}\right|_2 = \left.\frac{\rho_{\rm DR}}{\rho_{\rm R}}\right|_1 ~
\left({Y_{\phi_1} \over Y_{\phi_2}}\right)^{4/3}  \left({\Gamma_{\phi_2} \over \Gamma_{\phi_1}}\right)^{2/3}\,.
\ee
It is seen from (\ref{DMc}) and (\ref{DRc}) that the maximum dilution in the scenario with two moduli is achieved for $Y_{\phi_2} \simeq Y_{\phi_1}$ and $m_{\phi_2}\ll m_{\phi_1}$.\footnote{Note that $T_{{\rm R},2}/T_{{\rm R},1} \propto (\Gamma_{\phi_2}/\Gamma_{\phi_1})^{1/2} \propto (m_{\phi_2}/m_{\phi_1})^{3/2}$.} As pointed out earlier, in this case there is a single extended epoch of EMD consisting of stages 3 and 5 that are not separated by an intermediate RD phase.

\subsection{A string model with two epochs of modulus domination}
\label{Sec2moduli}

\subsubsection{The setup}

We now focus on a type IIB model which can allow for two epochs of modulus domination. This model shares the same features with the model discussed in Sec. \ref{Sec4} but it also gives rise to novel phenomenological properties. The Calabi-Yau volume now takes the form:
\be
\vo = \sqrt{\tau_{\rm vis}}\tau_{\rm big} - \tau_{\rm np}^{3/2}-\tau_{\rm inf}^{3/2}\,,
\ee
where again $\tau_{\rm inf}$ drives inflation and it is wrapped by a hidden sector D7 stack as in the model presented in Sec. \ref{Sec4}. However now the second blow-up mode, here denoted as $\tau_{\rm np}$, is just responsible for generating non-perturbative effects needed for moduli stabilization but it does not support the visible sector stack of D7 branes.\footnote{$\tau_{\rm np}$ instead supports a pure SYM hidden sector which generates gaugino condensation at a scale larger than the inflaton mass, so that the decay of $\sigma$ into heavy condensates on $\tau_{\rm np}$ is kinematically forbidden.} In fact, in this model the requirement to avoid dark radiation overproduction from the decay of the lightest modulus forces the visible D7 stack to be wrapped around the 4-cycle whose volume is controlled by $\tau_{\rm vis}$ \cite{Cicoli:2018cgu}.

In this case the 4D low-energy supergravity theory is determined by the following K\"ahler potential:
\be
K = -2\ln\left(\vo+\frac{\xi}{2 g_s^{3/2}}\right) + K_{g_s}\,,
\ee
where $K_{g_s}$ denotes string loop corrections \cite{Berg:2005ja,Berg:2007wt,Cicoli:2007xp} which have been shown to be $\vo$-suppressed with respect to the leading $\alpha'$ correction proportional to $\xi$ \cite{BBHL}. The superpotential instead looks like:
\be
W = W_0 + A_{\rm np}\,e^{-a_{\rm np} T_{\rm np}} + A_{\rm inf}\,e^{-a_{\rm inf} T_{\rm inf}}\,.
\ee
As in Sec. \ref{Sec4}, at leading order in $1/\vo \ll 1$, non-perturbative corrections to $W$ combined with $\alpha'$ corrections to $K$ produce an LVS minimum with 5 stabilized moduli: $\vo \simeq \sqrt{\tau_{\rm vis}}\tau_{\rm big}\sim e^{1/g_s}$, $\tau_{\rm np}\sim \tau_{\rm inf}\sim 1/g_s \sim \mathcal{O}(10)$ together with the 2 corresponding axions $c_{\rm np}$ and $c_{\rm inf}$. However at this level of approximation there are still 3 flat directions which can be parameterized by $\tau_{\rm vis}$, $c_{\rm vis}$, and $c_{\rm big}$. The visible sector modulus $\tau_{\rm vis}$ is fixed at subleading order by the string loop contribution to the K\"ahler potential $K_{g_s}$ at \cite{Cicoli:2008gp}:
\be
\tau_{\rm vis}\simeq g_s^{4/3}\, \lambda_{\rm loop}\, \vo^{2/3}\,,
\ee
where $\lambda_{\rm loop}$ is a tunable combination of the coefficients of $g_s$ corrections to $K$. Notice that the requirement of reproducing the observed value of the visible sector gauge coupling, $\alpha_{\rm vis}^{-1}=4\pi g_{\rm vis}^{-2}= \tau_{\rm vis}\sim \mc{O}(10-100)$, leads necessarily to an anisotropic shape of the extra dimensions since the exponentially large Calabi-Yau volume $\vo\simeq \sqrt{\tau_{\rm vis}}\tau_{\rm big}$ is now controlled by 2 cycles but with $\tau_{\rm big}\sim e^{1/g_s} \gg \tau_{\rm vis}\sim 1/g_s$. Finally, the two axions $c_{\rm vis}$ and $c_{\rm big}$ receive tiny masses due to additional $T_{\rm vis}$- and $T_{\rm big}$-dependent non-perturbative corrections to $W$. Thus both $c_{\rm vis}$ and $c_{\rm big}$ are in general ultra-light and play the role of hidden sector dark radiation.

\subsubsection{Moduli mass spectrum}

The mass spectrum of the relevant moduli around the minimum becomes:
\bea
m_\sigma^2 &\simeq& \kappa \,\epsilon^2 \left(\ln\epsilon\right)^2 \, M_{\rm P}^2\nn \\
m_{\phi_1}^2 &\simeq& \frac{\epsilon\,m_\sigma^2}{g_s^{3/2} W_0\, |\left(\ln\epsilon\right)^3|}\ll m_\sigma^2\quad\text{for}\quad \epsilon \ll 1 \nn \\
m_{\phi_2}^2 &\simeq& \frac{\epsilon^{1/3}\,g_s^{5/6} |\ln\epsilon|\,m_{\phi_1}^2}{W_0^{1/3}\sqrt{\lambda_{\rm loop}}} < m_{\phi_1}^2 \quad\text{for}\quad \epsilon, g_s \ll 1\nn \\
m^2_{a_{{\rm DR}_1}} &\sim& m^2_{a_{{\rm DR}_2}} \sim 0\,,
\label{ModMassSpectrum2}
\eea
where $\sigma$, $\phi_1$, $\phi_2$, $a_{{\rm DR}_1}$, and $a_{{\rm DR}_1}$ are the canonically normalized fields corresponding respectively to $\tau_{\rm inf}$, $\tau_{\rm big}$, $\tau_{\rm vis}$, $c_{\rm big}$, and $c_{\rm vis}$.

As in Sec. \ref{Sec4}, $\sigma$ plays the role of the inflaton. This field, when displaced from its minimum, becomes exponentially lighter than the Hubble constant during inflation which is set by the mass of $\phi_1$: $H \simeq m_{\phi_1}$. The 3 fields $\tau_{\rm np}$, $c_{\rm inf}$, and $c_{\rm np}$ are instead heavy spectator fields, while $\phi_1$ and $\phi_2$ get displaced from their minimum during inflation, and so give rise to 2 epochs of EMD. On the other hand, the 2 ultra-light axions $a_{{\rm DR}_1}$ and $a_{{\rm DR}_2}$ yield extra contributions to $N_{\rm eff}$. Moreover the gravitino mass and the soft terms are still given by (\ref{GravSoft}). Hence for $m_\chi \simeq m_0 \simeq M_{1/2}$, we conclude that DM cannot be reproduced from the decay of any of the 2 light moduli since:
\be
m_{\phi_2}^2 < m_{\phi_1}^2 \simeq \frac{\epsilon\,|\ln\epsilon|}{g_s^{3/2} W_0}\,m_\chi^2\ll m_\chi^2\quad\text{for}\quad\epsilon\ll 1\,.
\ee
Requiring $m_{\phi_2}\gtrsim\mathcal{O}(50)$ TeV in order to avoid any cosmological moduli problem together with $\tau_{\rm vis}\sim \mathcal{O}(100)$ in order to reproduce the observed value of the visible sector gauge coupling, corresponds to $1 \ll \vo \lesssim 5\times 10^7 - 10^9$ for $g_s\simeq 0.1$ and $1\lesssim W_0\lesssim 100$. Therefore DM is necessarily superheavy since $m_\chi \gtrsim 10^{11}$ GeV. Notice that values of the overall volume below $5\times 10^7 - 10^9$ are also required to match inflationary observables like the amplitude of primordial fluctuations \cite{Conlon:2005jm}.

\subsubsection{Moduli couplings and decay rates}

The configuration of the hidden sector D7-stack wrapped around $\tau_{\rm inf}$ is the same as the one described in Sec. \ref{HidSec}. Moreover, also the couplings of the inflaton $\sigma$ to hidden and visible gauge bosons are still given by (\ref{chidcvis}). Hence the ratio $\Gamma_{\sigma\to{\rm vis}}/\Gamma_\sigma$ is also still given by (\ref{ratiocvischid}) and the inflaton decay width $\Gamma_\sigma$ takes the same form as (\ref{InflatonWidth}).

On the other hand the light modulus $\phi_1$ decays mainly into visible sector gauge bosons with decay rate \cite{Cicoli:2010ha, Cicoli:2018cgu}:
\be
\Gamma_{\phi_1} \simeq \Gamma_{\phi_1\to{\rm vis}} = \gamma^2\frac{N_g}{96\pi}\frac{m_{\phi_1}^3}{M_{\rm P}^2}= \frac{\gamma^2}{8\pi}\frac{m_{\phi_1}^3}{M_{\rm P}^2}\quad\text{for}\quad N_g=12\,,
\ee
where $\gamma\geq 1$ is a microscopic parameter which depends on the gauge flux on the visible sector D7-stack (in particular $\gamma=1$ for a fluxless D7-stack while $\gamma>1$ for non-zero gauge fluxes) \cite{Cicoli:2018cgu}. The decay of $\phi_1$ produces also axionic dark radiation which is however suppressed for $\gamma >1$ and gets diluted by the decay of $\phi_2$. In what follows we shall therefore neglect $\Gamma_{\phi_1\to{\rm DR}}$.

The final modulus to decay is $\phi_2$ whose main decay channels are \cite{Cicoli:2018cgu}:
\bi
\item Dark radiation bulk axions:
\be
\Gamma_{\phi_2\to {\rm DR}} = \frac{5}{96\pi}\frac{m_{\phi_2}^3}{M_{\rm P}^2}\,,
\label{Gammaphi2DR}
\ee

\item Visible sector gauge bosons:
\be
\Gamma_{\phi_2\to{\rm vis}} = \gamma^2\frac{N_g}{48\pi}\frac{m_{\phi_2}^3}{M_{\rm P}^2} = \frac{\gamma^2}{4\pi}\frac{m_{\phi_2}^3}{M_{\rm P}^2}\,,
\label{Gammaphi2}
\ee
where we have set again $N_g=12$.
\ei

The amount of axionic dark radiation produced from $\phi_2$ decay is controlled by the underlying parameter $\gamma$:
\be
\Delta N_{\rm eff}\simeq 3\,\frac{\Gamma_{\phi_2\to{\rm DR}}}{\Gamma_{\phi_2\to{\rm vis}}} \simeq \frac{0.6}{\gamma^2}\,,
\ee
showing how for $\gamma\geq 1$ this model naturally satisfies present observational bounds since it yields $\Delta N_{\rm eff}\lesssim 0.6$.

The relevant quantities to compute the final DM abundance using (\ref{DMrelab2}) and (\ref{TR2}) can be derived from the decay widths (\ref{Gammaphi2DR}) and (\ref{Gammaphi2}) and read:
\be
\frac{\Gamma_{\phi_2}}{\Gamma_{\phi_2\to{\rm vis}}}= 1+\frac{5}{24\gamma^2}\,, \\
\ee
and:
\be
T_{{\rm R},2} \simeq 0.16 \gamma \left(1+\frac{5}{24\gamma^2}\right)^{1/4} m_{\phi_2}\sqrt{\frac{m_{\phi_2}}{M_{\rm P}}}\,.
\ee
Notice that the decay rates (\ref{Gammaphi2DR}) and (\ref{Gammaphi2}), together with the inflaton decay width (\ref{InflatonWidth}), also determine, via (\ref{DRrelab2}), the fractional energy density of DR.

\subsection{Inflationary observables and DM abundance}

As in Sec. \ref{Sec5}, we start analyzing the cosmology of the model with two moduli by presenting the expression for the total number of e-foldings:
\be
N_{\rm e} \simeq 57 + \frac{1}{4}\left[{\rm ln} ~ r - N_{\text{reh}} - N_{\phi_1} -N_{\phi_2} +\ln\left(\frac{\rho_{\sigma,\text{start}}}{\rho_{\sigma,\text{end}}} \right)\right], \nn
\ee
where $N_{\rm reh}$ is the duration of the reheating epoch after the end of inflation, while $N_{\phi_1}$ and $N_{\phi_2}$ denote respectively the number of e-foldings of the two EMD eras driven by the light moduli $\phi_1$ and $\phi_2$, which look like:
\be
N_{\phi_1} \simeq \frac{2}{3}\ln\left(Y_{\phi_1}^4 \frac{\Gamma_{\sigma}}{\Gamma_{\phi_1}}\right) ~ ~ , ~ ~  N_{\phi_2} \simeq \frac{2}{3} \ln \left(\left(\frac{Y_{\phi_2}}{Y_{\phi_1}} \right)^4 \frac{\Gamma_{\phi_1}}{\Gamma_{\phi_2}} \right). \nn
\ee
By rewriting $N_{\rm e}$ in terms of fundamental parameters, we obtain:
\be
N_{\rm e} \simeq 58.88 - \frac{1}{6}\ln \left(\frac{Y_{\phi_2}^4\vo^8}{W_0^5g_s^{3/2} \left|\ln \epsilon \right|^{3/2}} \right),
\ee
where we have set $\gamma=1$. Notice that the total number of e-foldings does not depend on the initial misalignment value of the first modulus.

As in the single modulus scenario, we obtain $W_0$ as a function of $\vo$ by combining two constraints coming from the amplitude of the primordial scalar fluctuations and the geometrical requirement to have the volume of blow-up modes hierarchically smaller than the overall internal volume. Following the same procedure as in Sec. \ref{Sec5}, we then extract the value of $\vo$ from matching the observed DM abundance. Finally, this value of the volume fixes the DM mass for every combination of the underlying parameters. Interestingly, all data points correspond to a DM mass in the same range as in the single modulus case, $m_\chi \simeq 10^{10}$-$10^{11}$GeV, with a bias towards smaller values (65\% of the data points result in $m_\chi \simeq 10^{10}$GeV). 

\begin{figure}[ht!]
    \centering
     \includegraphics[trim=0cm 0cm 0cm 0cm, clip=true, width=0.48\textwidth]{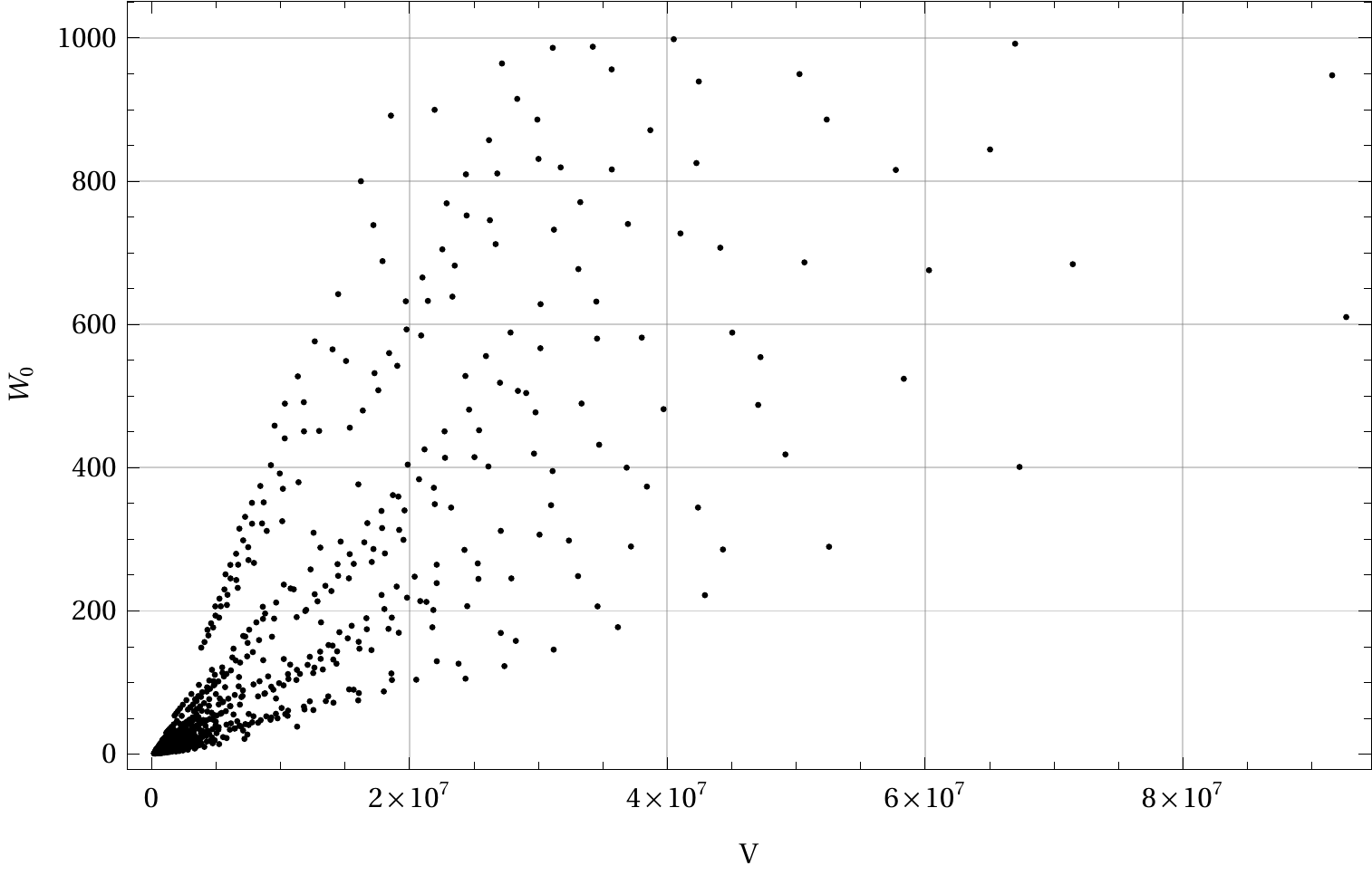}
    \caption{Points in the $(W_0,\vo)$ plane which match the observed amplitude of the density perturbations and DM abundance.}
\label{FigApp}
\end{figure}

In Fig. \ref{FigApp} we present the points in the ($W_0$, $\vo$) plane which satisfy all our theoretical and phenomenological conditions. Note that 65\% of the initial parameter set reproduces the correct DM abundance. Moreover we were not able to obtain values of $n_{\rm s}$ within the $2\sigma$ range, as each point in Fig. \ref{FigApp} corresponds to $n_{\rm s}$ at the lower end of the $3\sigma$ range. The qualitative behavior of the underlying parameters $g_s$, $Y_{\phi_2}$, and $\lambda$ is the same as in the single modulus case. Tab. \ref{Tab2} shows a representative choice of the parameters used to perform a numerical study of the full cosmological evolution in this scenario.

\begin{table}[h]
\begin{center}
 \begin{tabular}{||c || c | c ||} 
\hline
$W_0$ & 20.4 \\
\hline 
$\vo$ &   $3 \times 10^6$  \\
 \hline
$N_{\rm e}$ & 44.4 \\
\hline
$N_{\text{reh}}$ & 4.4 \\
\hline
$N_{\phi_1}$ & 14.8 \\
\hline
$N_{\phi_2}$ & 3.9 \\
\hline
$n_{\rm s}$ & 0.9550\\
\hline
$m_{\sigma}$ & $1.1 \times 10^{13} \text{GeV}$ \\
\hline
$m_{\phi_1}$ & $7.7 \times 10^8\text{GeV}$\\
\hline
$m_{\phi_2}$ & $8.3 \times 10^7\text{GeV}$\\
\hline
$m_{3/2}$ & $8.9 \times 10^{11}\text{GeV}$\\
\hline
$m_{\chi}$ & $7.3 \times 10^{10} \text{GeV}$\\
\hline
$c_{\text{hid}}$ & 332.0\\
\hline
\end{tabular}
\caption{Microscopic parameters $W_0$ and $\vo$, the resulting e-folding numbers, $n_{\rm s}$, mass scales, and inflaton coupling to hidden degrees of freedom $c_{\rm hid}$ at a benchmark point that gives the right amplitude of the density perturbations and the correct DM abundance. The input parameters are $g_s = 0.1$, $Y_{\phi_1} = Y_{\phi_2} = 0.01$, $\lambda = 10^3$, $N_g^{\rm hid}=12$, and $\gamma = 1$.}
\label{Tab2}
\end{center}
\end{table}

\vspace{-0.55cm}
\subsection{Numerical analysis of cosmological evolution}
\label{NumCosmEv2}
% \vspace{-0.1cm}

We obtain the numerical evolution of energy densities for the scenario with two moduli.  As in Sec. \ref{numerical_evolution}, we begin the evolution at \(H \simeq H_{\rm inf}\) with both \(\sigma\) and \(\phi_1\) oscillating. Though $\phi_2$ begins oscillating shortly after this time, its energy density is subdominant and its initial evolution can be approximated as a matter component without altering the overall evolution. The other energy density components are again highly suppressed initially. The Boltzmann equations for this scenario are (neglecting the tiny production of DR from the decay of $\phi_1$):
\begin{align}
    & \frac{d\rho_\sigma}{dt} + 3H\rho_\sigma = - \Gamma_\sigma\,\rho_\sigma \, ,
    \label{boltz2_inf}
\\
    & \frac{d\rho_{\phi_1}}{dt} + 3H\rho_{\phi_1} = -\Gamma_{\phi_1}\,\rho_{\phi_1} \, ,
    \label{boltz2_phi1}
\\
    & \frac{d\rho_{\phi_2}}{dt} + 3H\rho_{\phi_2} = - \Gamma_{\phi_2}\,\rho_{\phi_2} \, ,
    \label{boltz2_phi2}
\\
    & \frac{d\rho_{\rm DR}}{dt} + 4H\rho_{\rm DR} = \Gamma_{\sigma\to{\rm DR}} \rho_\sigma + \Gamma_{\rm\phi_2 \rightarrow DR}\rho_{\phi_2} \, ,
    \label{boltz2_DR}
\\
    & \frac{d\rho_{\rm R}}{dt} + 4H\rho_{\rm R} = \Gamma_{\rm \sigma \rightarrow vis}\rho_\sigma + \Gamma_{\phi_1\to{\rm vis}}\rho_{\phi_1} + \Gamma_{\phi_2\to{\rm vis}}\rho_{\phi_2} \, ,
    \label{boltz2_R}
\\
       & \frac{dn_\chi}{dt} + 3Hn_\chi = {\rm Br}_\chi\Gamma_\sigma\left(\frac{\rho_\sigma}{m_\sigma}\right)
       + \left<\sigma_{\rm ann}v\right>\left(n_{\rm\chi,eq}^2 - n_\chi^2\right) \, .
       \label{boltz2_chi}
\end{align}

\begin{figure}[htb]
    \centering
    \includegraphics[trim=0cm 0cm 0cm 0.3cm, clip=true, width=0.48\textwidth]{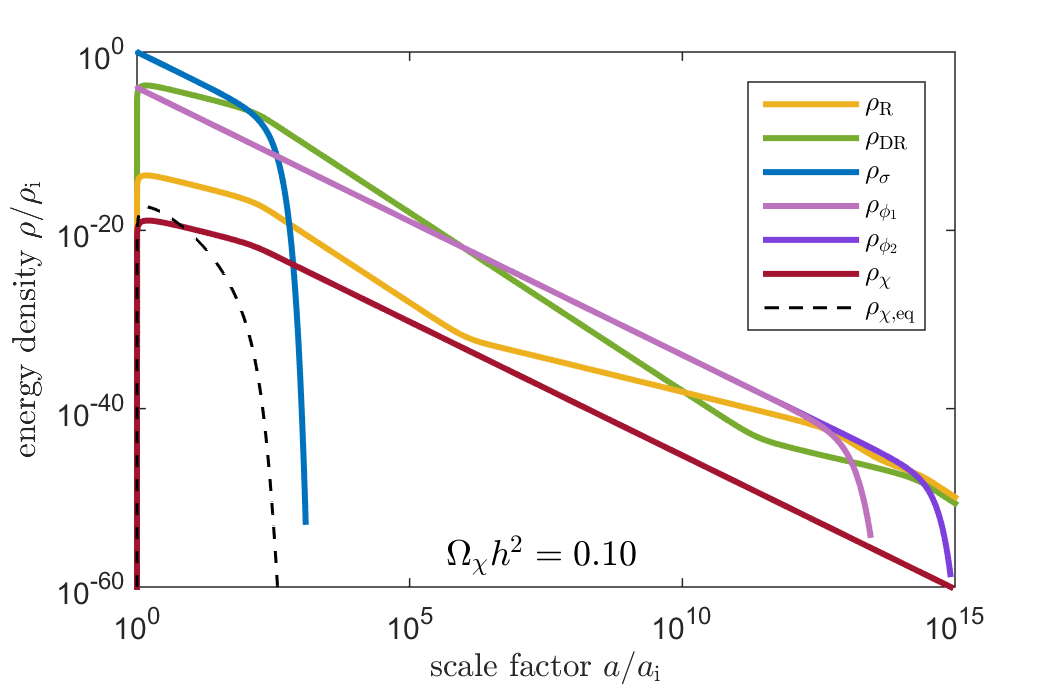}
    \caption{Energy density evolution of the various components as functions of scale factor in the two-moduli scenario for the benchmark point in Tab. \ref{Tab2}.}
    \label{fig:2mod_rho}
\end{figure}

\begin{figure}[htb]
    \centering
      \includegraphics[trim=0cm 0cm 0cm 0.3cm, clip=true, width=0.48\textwidth]{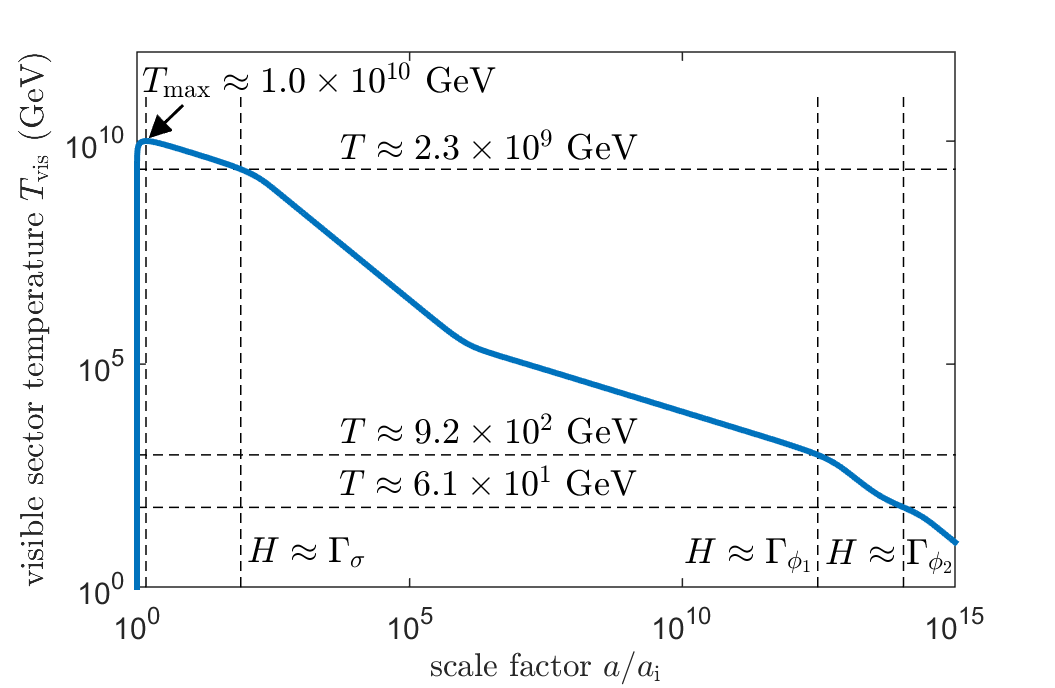}
    \caption{Visible sector temperature as a function of scale factor in the scenario with two moduli.}
    \label{fig:2mod_HT}
\end{figure}

A sample numerical solution is shown in Fig. \ref{fig:2mod_rho} for the becnhmark point in Tab. \ref{Tab2}. The cosmological evolution is very similar to the one-modulus case. The effect of the second, lighter, modulus is to extend the EMD period to lower temperatures. Because the two moduli start with equal misalignments, we have a single extended period of EMD with a brief period of substantial radiation while the lighter modulus dominates, instead of two EMD periods separated by a RD phase. Fig. \ref{fig:2mod_HT} shows the visible sector temperature as a function of the scale factor, where we have taken the temperature dependence of the relativistic degrees of freedom into account.

\end{document}